\documentclass{article}
\usepackage{arxiv}

\usepackage[utf8]{inputenc} 
\usepackage[T1]{fontenc}    
\usepackage{hyperref}       
\usepackage{url}            
\usepackage{booktabs}       
\usepackage{amsfonts}       
\usepackage{nicefrac}       
\usepackage{microtype}      
\usepackage{lipsum}		
\usepackage{graphicx}
\usepackage{doi}

\usepackage{cite}
\usepackage{amsmath,amssymb,amsfonts}
\usepackage{algorithmic}
\usepackage{graphicx,color}
\usepackage{textcomp}
\usepackage{xcolor}
\usepackage{hyperref}
\hypersetup{hidelinks=true}
\UseRawInputEncoding
\usepackage{float}
\usepackage[linesnumbered,ruled,vlined]{algorithm2e}
\usepackage{soul}
\SetCommentSty{mycommfont}
\SetKwInput{KwInput}{Input}                
\SetKwInput{KwOutput}{Output}              

\usepackage{mathrsfs} 
\usepackage{bm}
\usepackage{subcaption}
\usepackage{graphicx}
\usepackage{comment}

\title{Energy-Efficient On-Board Radio Resource Management for Satellite Communications via Neuromorphic Computing}

\author{ \href{https://orcid.org/0000-0002-2280-4689}{\includegraphics[scale=0.06]{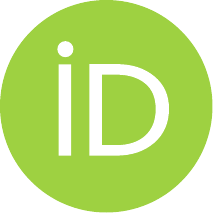}\hspace{1mm}Flor Ortiz}\thanks{The first and second author contributed equally to this work. Corresponding author: Flor Ortiz (email: flor.ortiz@uni.lu).} \\
	Interdisciplinary Centre for\\ Security, Reliability, and Trust\\ (SnT),
 University of Luxembourg\\
	\And
	{\hspace{1mm}Nicolas Skatchkovsky}\\
	Francis Crick Institute, \\
        London, United Kingdom 	
 \And
	{\hspace{1mm}Eva Lagunas} \\
	Interdisciplinary Centre for\\ Security, Reliability, and Trust\\ (SnT),
 University of Luxembourg\\
  \And
	{\hspace{1mm}Wallace A. Martins} \\
	Interdisciplinary Centre for\\ Security, Reliability, and Trust\\ (SnT),
 University of Luxembourg\\
  \And
	{\hspace{1mm}Geoffrey Eappen} \\
	Interdisciplinary Centre for\\ Security, Reliability, and Trust\\ (SnT),
 University of Luxembourg\\
  \And
	{\hspace{1mm}Saed Daoud} \\
	Interdisciplinary Centre for\\ Security, Reliability, and Trust\\ (SnT),
 University of Luxembourg\\
 \And
	{\hspace{1mm}Osvaldo Simeone} \\
	Department of Engineering\\King's College London,\\ United Kingdom
 \And
	{\hspace{1mm}Bipin Rajendran} \\
	Department of Engineering\\King's College London,\\ United Kingdom
 \And
	{\hspace{1mm}Symeon Chatzinotas} \\
	Interdisciplinary Centre for\\ Security, Reliability, and Trust\\ (SnT),
 University of Luxembourg\\
}

\renewcommand{\shorttitle}{\textit{arXiv} Template}

\hypersetup{
pdftitle={A template for the arxiv style},
pdfsubject={q-bio.NC, q-bio.QM},
pdfauthor={David S.~Hippocampus, Elias D.~Striatum},
pdfkeywords={First keyword, Second keyword, More},
}

\begin{document}
\maketitle

\begin{abstract}
The latest satellite communication (SatCom) missions are characterized by a fully reconfigurable on-board software-defined payload, capable of adapting radio resources to the temporal and spatial variations of the system traffic. As pure optimization-based solutions have shown to be computationally tedious and to lack flexibility, machine learning (ML)-based methods have emerged as promising alternatives. We investigate the application of energy-efficient brain-inspired ML models for on-board radio resource management. Apart from software simulation, we report extensive experimental results leveraging the recently released Intel Loihi 2 chip. To benchmark the performance of the proposed model, we implement conventional convolutional neural networks (CNN) on a Xilinx Versal VCK5000, and provide a detailed comparison of accuracy, precision, recall, and energy efficiency for different traffic demands. Most notably, for relevant workloads, spiking neural networks (SNNs) implemented on Loihi 2 yield higher accuracy, while reducing power consumption by more than 100$\times$ as compared to the CNN-based reference platform. Our findings point to the significant potential of neuromorphic computing and SNNs in supporting on-board SatCom operations, paving the way for enhanced efficiency and sustainability in future SatCom systems.
\end{abstract}


\section{Introduction}

\subsection{Context and Motivation}
Satellite Communications (SatCom) have become increasingly important in recent years due to the surge in global connectivity demands. With the integration of terrestrial systems like 6G and the pressing need to reduce the digital divide and obtain ubiquitous coverage, SatCom plays a vital role in bridging the communication gap worldwide~\cite{Wang2023OnThings,Kodheli2021SatelliteChallenges, Ortiz-Gomez2017MethodLink}. However, the growing traffic demand in SatCom systems presents significant challenges in effectively managing the allocation of radio resources to meet Quality-of-Service (QoS) requirements while minimizing resource utilization ~\cite{Kisseleff2021RadioSystems,Ortiz-Gomez2020SupervisedSystems}.

Conventional SatCom systems typically employ static multi-beam configurations with fixed bandwidth and power allocations. These systems are incapable of adapting to the dynamic nature of traffic demands. As a result, resources may be wasted, while user demands may be left unfulfilled. Recognizing the temporal and spatial variations in demand, software-defined payloads have emerged, offering unprecedented flexibility and adaptability in radio resource management (RRM) for SatCom  ~\cite{Kisseleff2021RadioSystems}.

Software-defined payloads have revolutionized the SatCom landscape by providing fully reconfigurable systems capable of dynamically allocating power and bandwidth resources. The advent of these payloads has driven the need for effective RRM techniques to optimize resource allocation and ensure efficient utilization. While traditional optimization-based solutions have been explored for RRM,  they are often computationally cumbersome and lack the required flexibility to address the diverse and dynamic traffic patterns encountered in SatCom systems~\cite{8742502}.

In recent years, machine learning (ML) algorithms have emerged as a promising alternative to conventional optimization approaches for RRM in SatCom~p{9374638,9177044}. ML-based solutions not only offer the potential to adaptively learn and predict traffic patterns, but also contribute to accelerate complex RRM algorithmics and bring adaptation and flexibility to static optimization solutions. However, implementing ML algorithms on board may be problematic due to the potentially high energy budgets incompatible with satellites' available resources. 

\subsection{Contributions}

To address this problem, this paper investigates neuromorphic computing as an alternative to conventional neural network-based platforms to enhance efficiency and sustainability of on-board SatCom operations~\cite{ortiz2022towards} (see Fig. \ref{fig:flexible}). Neuromorphic processors (NPs) represent a new class of computing devices inspired by the human brain's architecture and computational principles. They offer unique advantages in terms of low power consumption, high parallelism, and real-time processing capabilities ~\cite{Park2020NeuromorphicOutlook}. Leveraging these benefits, NPs hold significant promise for enabling energy-efficient RRM on board satellites.

To provide empirical evidence and benchmark the performance of NPs, we conduct experiments using Intel's cutting-edge neuromorphic processor, Loihi 2. Benchmarking is carried out by comparing the accuracy, precision, recall, and energy efficiency of NPs with conventional platforms such as Xilinx Versal VCK5000 under varying traffic demands for the problem of RRM. Particularly noteworthy are the results obtained using spiking neural networks (SNNs) implemented on the Loihi processor, demonstrating superior accuracy while reducing power consumption by more than a factor of 100 compared to the reference platform based on convolutional neural networks (CNNs).

\subsection{Related Work}

\subsubsection{Machine Learning for RRM in SatCom}
The efficient implementation of on-board RRM is crucial for optimizing performance and ensuring seamless connectivity in SatCom systems. In recent years, ML techniques have gained significant attention for resource management in various SatCom scenarios. A study by~\cite{Ortiz-Gomez2022MachineSystems} focused on ML-based resource management in multi-beam geostationary Earth orbit (GEO) satellite systems. The authors analyze different ML techniques applied to systems with power, bandwidth, and/or beamwidth flexibility and systems with beam hopping capabilities. Furthermore, reference~\cite{Wang2021Dual-DNNSystems} proposes a combined learning and optimization approach to address a mixed-integer convex programming problem (MICP) in satellite RRM. The problem is decomposed into classification-like tasks and power control optimization, respectively solved by dual deep neural networks (DNNs) and convex optimization. 

Another notable work by Deng et al.~\cite{Deng2020TheLearning} focuses on resource management in next-generation heterogeneous satellite networks (HSNs),  and introduces an innovative framework that encourages cooperation among independent satellite systems to maximize resource utilization. Deep reinforcement learning (DRL) optimizes resource allocation and supports intercommunication between different satellite systems. In a related study,  Ferreira et al.~\cite{Ferreira2018MultiobjectiveEnsembles} proposed a feasible solution for real-time, single-channel resource allocation problems using DRL.  Their study discretized resources before allocation, which may not be optimal for continuous resources like satellite power. Luis et al. 
\cite{Luis2019DeepSatellites} addressed this issue by exploring a DRL architecture with constant, stateful action spaces for energy allocation, avoiding the need for discretization. Liu et al.~\cite{Liu2020AG-DPA:Systems, Liu2018DeepSystems} presented a DRL-based dynamic channel allocation algorithm (DRL-DCA) for multi-beam satellite systems, achieving lower blocking probabilities than traditional algorithms. Finally, Liao et al.~\cite{Liao2020DistributedAllocation} also introduced a game model and a DRL-based bandwidth allocation framework for satellite communication scenarios, dynamically allocating bandwidth in each beam. The proposed method effectively handles time-varying traffic and large-scale communication, albeit limited to managing a single resource on the satellite.

\subsubsection{Neuromorphic Computing for Communications} 

The application of neuromorphic processors to communications has gained significant interest in recent years as a low-power alternative to traditional systems. The work \cite{Skatchkovsky2021SpikingCommunications} explores the application of SNNs for learning and inference in battery-powered devices connected over bandwidth-constrained channels. It summarizes activity on  federated learning for distributed training of SNNs \cite{skatchkovsky2020federated} and the integration of neuromorphic sensing, SNNs, and pulse radio technologies for low-power remote inference \cite{Skatchkovsky2020End-to-EndIntelligence}. The recent references~\cite{Chen2022NeuromorphicCommunications,Chen2023NeuromorphicInference} generalized wireless neuromorphic communications to multi-access channels and introduced the concept of neuromorphic integrated sensing and communications (N-ISAC). N-ISAC utilizes a common impulse radio waveform for transmitting digital information and detecting radar targets, employing an SNN for decoding and target detection. The optimization of SNN operation balances data communications and radar sensing performance metrics, showcasing the synergies and trade-offs between these applications.

\subsubsection{Neuromorphic Computing for SatCom}

While the field of neuromorphic learning for communications is still in its infancy, there is a growing interest in its applications within the space sector due to its high energy efficiency and promising performance~\cite{Kucik2021InvestigatingClassification}. Both academia and industry sectors have significantly advanced in developing non-space applications for neuromorphic computing, which the space industry can potentially leverage as ``spin-in" technology.

Recent research initiatives further highlight the increasing interest in exploring the potential benefits of neuromorphic computing. NASA's launch of TechEdSat-13, equipped with Intel's Loihi neuromorphic processor, into low Earth orbit (LEO) in 2022 is a testament to their commitment to testing new capabilities for future AI science and engineering applications in space~\cite{NEuromorphicNASA}. Additionally, the European Space Agency (ESA) initiated an ARTES Future Preparation activity in 2021~\cite{NeuroSatTIA}, focusing on exploring the use of neuromorphic computing for SatCom systems, underscoring the significance of this technology within the space sector.

Our previous work, published in 2022, highlights the potential use cases and applications of neuromorphic processors for SatCom~\cite{ortiz2022towards}. Neuromorphic processors, such as Loihi~\cite{Davies2018Loihi:Learning}, are, in fact, particularly well-suited for processing sparse time series data since their energy requirements are proportional to the number of emitted spikes, which occur only in the presence of relevant events.

\subsection{Organization}
The document is structured as follows. Section~\ref{sec:system} provides a detailed overview of the system model, covering the problem definition and the traffic demand model. Section~III discusses our benchmark approach based on CNNs. In Section~IV, we describe the approach based on neuromorphic computing. Section~V discusses the implementation of both models on the respective hardware. Section~VI presents experimental results, showing the superiority of the neuromorphic computing approach for practical workloads. Finally, Section~VII offers a detailed analysis of the results and draws some conclusions.

\section{System Model}\label{sec:system}
\begin{figure*}[ht]
\centering
\includegraphics[width=0.8\linewidth]{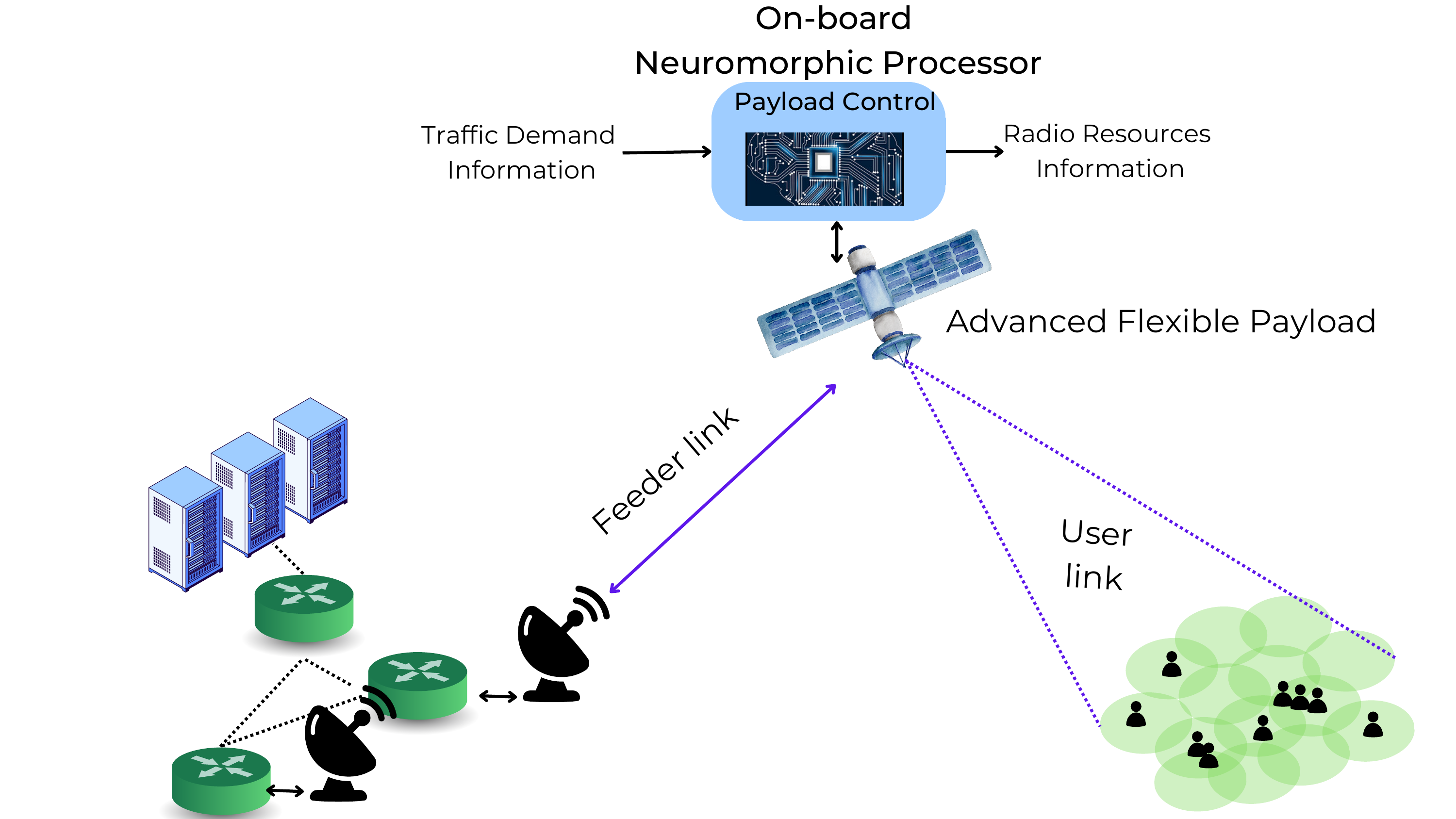}
\caption{On-board resource management for flexible payload in a multibeam satellite}
\label{fig:flexible}
\end{figure*}

As illustrated in Fig. \ref{fig:flexible}, the system architecture comprises a flexible, software-defined radio (SDR) satellite payload that allows for adjustable bandwidth and power allocation to each beam. The management of communication resources is dynamically adapted in response to changes in traffic demand. Specifically, adopting a data-driven ML-based solution, we focus on designing ML models that take the traffic demand over the service area as input, and output an optimized resource configuration. After offline training, the ML payload controller can be deployed on board the satellite for real-time inference. This data-driven architecture offers the advantage of reduced processing times~\cite{aerospace10020101}. Table~\ref{tab:Onboard_flexiblepayload} provides an overview of the main characteristics of the setting under study.

\begin{table}[ht]
\caption{On-board RRM for flexible payload scenario}\label{tab:Onboard_flexiblepayload}
\begin{tabular}{|l|l|}
\hline
Scenario    & On-board RRM for flexible payload                                                                                  \\ \hline
System architecture   & SDR payload                                                                                                      \\ \hline
Air interface         & Any air interface supporting multicarrier                                                                                \\ \hline
AI-based technique(s) & Supervised learning: Classification                                                                       \\ \hline
Input / Output        & \begin{tabular}[c]{@{}l@{}}Input: Geographical traffic demand distribution \\  Output: Power and bandwidth configuration\end{tabular} \\ \hline
\end{tabular}
\end{table}

In more detail, we consider a GEO high-throughput satellite system consisting of a single multibeam satellite that provides coverage to a wide region of Earth through $B$ spot-beams. We focus on the forward link, consisting of uplink feeder and downlink user links, and we assume a total of $K$ single-antenna user terminals (UTs) distributed across the satellite coverage area. The considered payload can flexibly manage  power using a traveling-wave tube amplifier (TWTA) with adaptive input back-off (IBO). Furthermore, it manages spectrum utilization via a channelizer on board that separates signals into frequency blocks and rearranges them to achieve flexible bandwidth allocations and to avoid co-channel beam interference, frequency reuse is assumed~\cite{Ortiz-Gomez2022MachineSystems}.

\subsection{Problem Statement}
\label{problem_statement_label}
The heterogeneous distribution of traffic demands across the satellite beams and over the satellite's lifetime motivates the use of dynamic RRM~\cite{Kisseleff2021RadioSystems}. 
The objective of RRM is to efficiently allocate available bandwidth and power  resources to minimize the discrepancy between the offered capacity $C_{\tau}^b$ and the requested capacity $R_{\tau}^b$ on each  $b$-th beam during any time slot $\tau$.

The offered capacity $C_{\tau}^b$ [bps] in the $b$-th beam during time slot $\tau$ can be calculated as
\begin{equation}C_{\tau}^b= {\rm W}_{\tau}^{b}\cdot{\rm \kappa}_{\tau}^b,\label{offcap_eq}\end{equation} 
where ${\rm \kappa}_{\tau}^b$ [bps/Hz] is the  spectral efficiency of the selected modulation and coding (ModCod) scheme, and ${\rm W}_{\tau}^{b}$ [Hz] denotes the bandwidth allocated to $b$-{th} beam. 

The spectral efficiency ${\rm \kappa}_{\tau}^b$ depends on the the carrier-to-interference-plus-noise ratio (CINR) $\gamma^b_{\tau}$ of the $b$-{th} beam at slot $\tau$. Accordingly, we write it as a generic function $\kappa_b = f(\gamma_b)$, where $f(\cdot)$ denotes the mapping between CINR and the selected ModCod scheme, which can be found in standards such as DVB-S2X \cite{ETSIEngineering360}. 

The CINR $\gamma^b_{\tau}$  in turn depends on the power $P_{\tau}^b$ [W] and bandwidth ${\rm W}_{\tau}^{b}$ [Hz] allocated to the $b$-{th} beam. In particular,  we have 
\begin{equation}
\gamma_{N,\tau}^b = \frac{P_{\tau}^b|h^b|^2}{I_{\tau}^b+N_{0} {\rm W}_{\tau}^{b}},
\end{equation} where $N_{0}$ is the power spectral density of noise, and  $|h^b|^2$ represents the channel gain for beam $b$. For the latter, we  use the standard model
\begin{equation}
|h_\tau^b|^2 = \frac{G_{\text{SAT}} \left(\theta^{b}_\tau\right) G_{\text{RX,max}}}{\left(4 \pi D_{\tau}^b/{\lambda}\right)^{2}L_{\tau}^b},
\label{eq}
\end{equation}
where $D_\tau^b$ is the distance between the satellite and $b$-{th} beam center on the ground;  $\lambda$ indicates the wavelength; $L_{\tau}^b$ denotes the shadowing and atmospheric gas losses; and the satellite off-boresight transmit angle to the different beams is  $\theta_{\tau}^{b}$. The variables  $G_{\text{SAT}}\left(\theta\right)$ and $G_{\text{RX,max}}$ denote the satellite antenna gain towards a specific off-boresight angle $\theta$ and the user terminal receive antenna gain, respectively. The user terminal antenna is assumed to point towards the GEO satellite, and therefore, the received antenna gain is fixed and equal to the maximum supported by the receiver's antenna.

Inspired by \cite{Ortiz-Gomez2020SupervisedSystems}, our goal is to use minimal power and spectral bandwidth to match the offered capacity given in \eqref{offcap_eq} to the aggregated traffic demand $R_{\tau}^b$ [bps] of each beam $b$ at each particular time instant $\tau$.  In particular, focusing on a specific time slot $\tau$, the objective function can be formulated as
\begin{equation}
\label{eq:obj}
U(P_\tau,W_\tau)=\beta_0\sum_{b=1}^{B}|C_{\tau}^b - R_{\tau}^b|- \beta_1\sum_{b=1}^{B}P_{\tau}^b  - \beta_2\sum_{b=1}^{B}{\rm W}_{\tau}^{b},
\end{equation} 
where $P_\tau$ and $W_\tau$ denote the set of power and bandwidth allocation variables $\{P^b_\tau\}$ and $\{\rm W^b_\tau\}$, respectively. The first term in (\ref{eq:obj}) denotes the mismatch between the offered capacity $C_{\tau}^b$ and the requested capacity $R_{\tau}^b$. The second term facilitates the minimization of the total transmit power of the satellite system, i.e. $\sum_{b=1}^{B}P_{\tau}^b$. Finally, the third term prioritizes solutions in which the total bandwidth $\sum_{b=1}^{B}{\rm W}_{\tau}^{b}$ allocated to the beams is minimized. The hyperparameters $\beta_0\geq 0$, $\beta_1\geq 0$ and $\beta_2\geq 0$ determine the relative weights of these criteria. 

The objective function (\ref{eq:obj}) is minimized under total bandwidth constraints and total system power constraint as \begin{equation}
\begin{array}{ll}
\underset{P_\tau,W_\tau}{\operatorname{minimize}} &U(P_\tau,W_\tau) \\ 
\text { s.t. } &  \sum_{b=1}^{B} P_{\tau}^b \leq P_{\max} \\
&  \sum_{b=1}^{B} {\rm W}_{\tau}^{b} \leq {\rm W}_{\max}.\\\
\end{array}\label{eq:1}  
\end{equation} 
The objective (\ref{eq:obj}) includes terms supporting  power and bandwidth minimization so as to potentially reduce the power and bandwidth consumption beyond the upper bounds imposed by the constraints in (\ref{eq:1}).


We assume that power and bandwidth variables $P^b_\tau$ and $W^b_\tau$ are to be selected within discrete sets  $\mathcal{P}_\tau = \{ P_1, P_2, \ldots, P_N \}$ and $\mathcal{W}_\tau = \{ W_1, W_2, \ldots, W_M \}$ of feasible solutions, imposing the additional constraints $P_{\tau}^b \in \mathcal{P}_\tau$ and ${\rm W}_{\tau}^{b} \in \mathcal{W}_\tau$. 
As a result, problem (\ref{eq:1}) is of combinatorial nature, as it involves the selection of variables from a finite set of options, and its complexity scales exponentially with the number of beams. Since the aggregated beam demands $R_{\tau}^b$ generally change over time, problem (\ref{eq:1}) needs to be solved anytime that there is a relevant change in the traffic profile, resulting in an update on the power and bandwidth assignment. 



\subsection{Traffic Model} \label{sec:traffic}
\begin{figure*}
  \centering
  \begin{subfigure}{0.49\textwidth}
    \includegraphics[width=\textwidth]{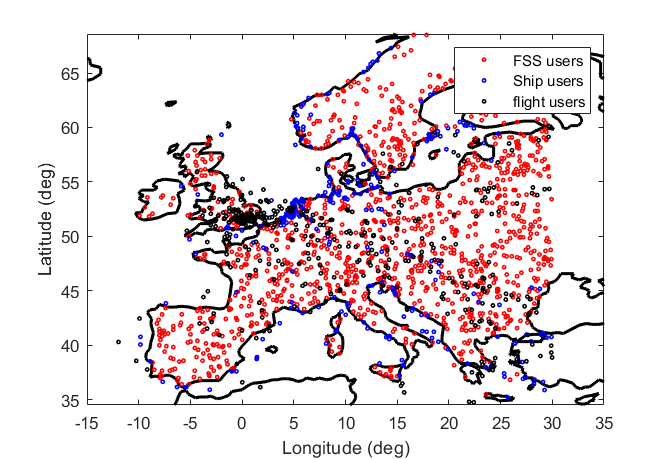}
    \caption{Traffic demand at 2 am}
    \label{fig:sub1}
  \end{subfigure}
  \hfill
  \begin{subfigure}{0.49\textwidth}
    \includegraphics[width=\textwidth]{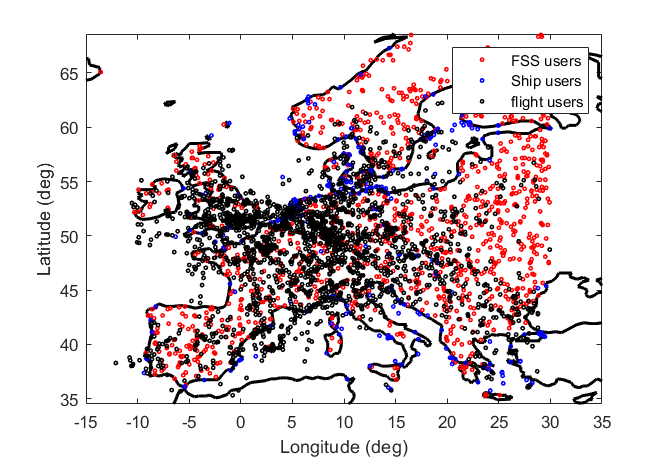}
    \caption{Traffic demand at 10 am}
    \label{fig:sub2}
  \end{subfigure}
  \caption{The traffic demand model in Europe depends on population density, aerial, and maritime density. }
  \label{fig:traffic}
\end{figure*}

In order to model the requested capacity $R_{\tau}^b$,
we make use of the traffic simulator developed by University of Luxembourg and presented in ~\cite{Traffic}, which integrates population, aeronautical, and maritime data to accurately model traffic demand and distribution within a specific service area. As depicted in Fig.~\ref{fig:traffic}, first, the population dataset considers the distribution of broadband fixed satellite service (FSS) terminals, encompassing the fundamental spatial patterns of FSS traffic. The population data has been obtained from the NASA Socioeconomic Data and Applications and Data Center (SEDAC) population density database~\cite{NASA_population}. 
Secondly, the simulator incorporates real variations in aeronautical traffic by utilizing data extracted from an anonymous and unfiltered flight tracking source~\cite{Traffic,ADS-B}. This inclusion enables the examination of the impact of flight volume on the geographical density of traffic at different time intervals, thereby ensuring an accurate representation of the spatio-temporal distribution of aeronautical traffic.

Lastly, the maritime dataset accounts for the potential demand for satellite connectivity through ship communications, which exhibit significant changes over time and location. To capture this, the simulator employs a dataset obtained from vessel traffic services (VTS), comprising vessel positions and maritime traffic detected by the global automatic identification system (AIS)~\cite{Traffic,AIS}.

To prepare the collected datasets for analysis, a pre-processing unit within the traffic simulator handles tasks such as eliminating redundant and conflicting traffic records, resolving missing information, and extracting user positions. UTs are categorized and assigned to their respective service beams based on the geographic longitudes and latitudes. The simulator also considers the limited use of FSS in large urban areas, recognizing the prevalence of alternative broadband technologies in such regions.

For modeling daily hourly traffic demands, the aeronautical data traces are pre-processed by collecting and gathering flight data for one hour. Similarly, in capturing the temporal maritime traffic demand, the maritime data is analyzed, taking into account the position of each ship's first appearance within the covered area during each hour. This approach allows for a reasonable estimation of the current and anticipated demand within an hour~\cite{Traffic}.

\section{Conventional ML Benchmark}

\begin{figure*} [ht]
\centering
\includegraphics[width=0.95\linewidth]{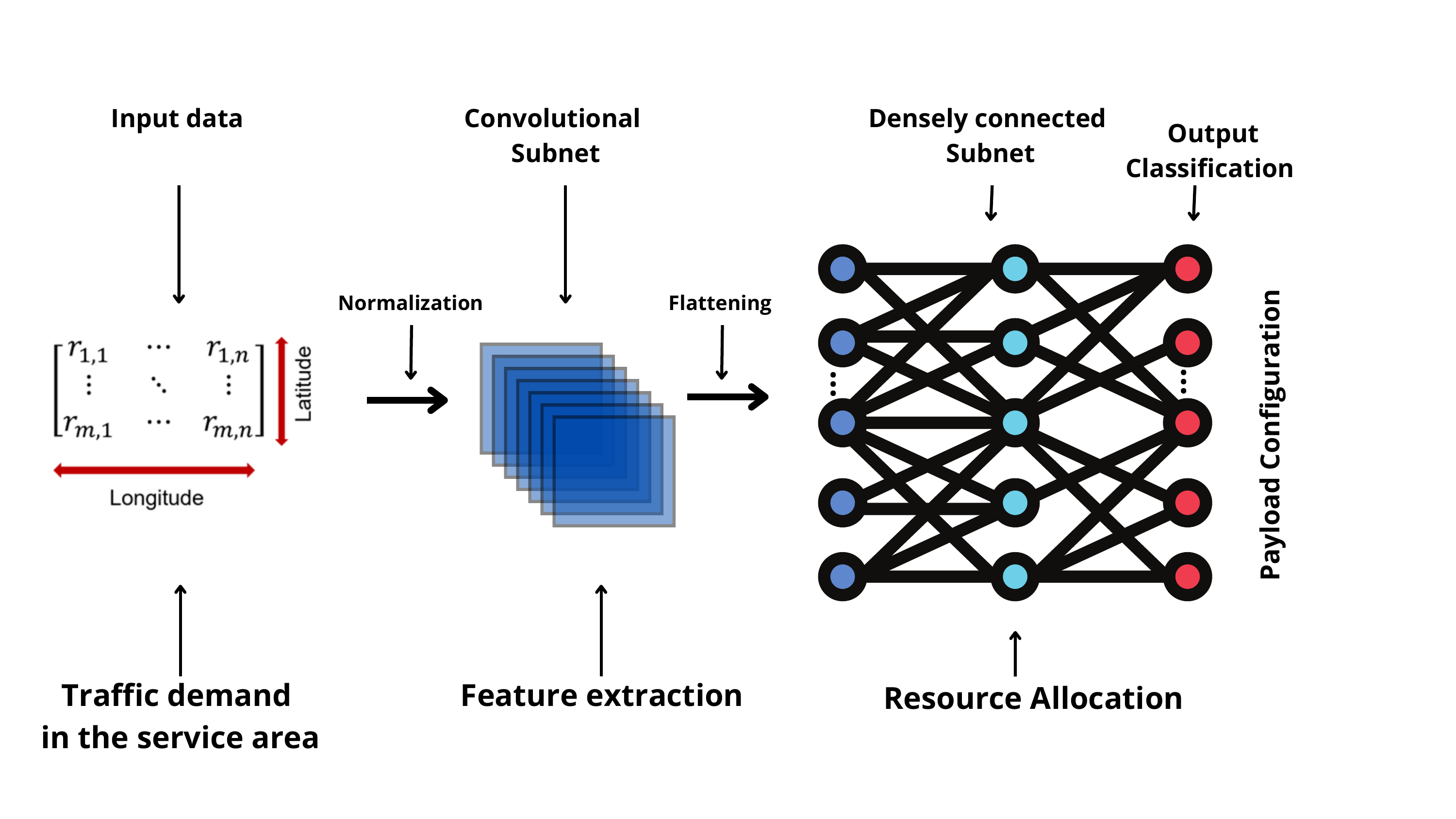}
\caption{Benchmark approach based on convolutional neural networks (CNNs) for on-board RRM.}
\label{fig:CNNflexible}
\end{figure*}

In this section, we describe for reference the CNN architecture that will be used as benchmark solution when implemented on a Xilinx Versal VCK5000.

\subsection{Architecture}
 
A CNN architecture for classification task is introduced to solve the problem \eqref{eq:1}. The CNN's detailed design is illustrated in Figure \ref{fig:CNNflexible} and will be detailed in the following. In the input layer of the CNN, we have the per-beam demand requests formatted in a matrix form $\bm{R}_{\tau}$, which  represents the traffic demand at each geographic location in the service area as\begin{align}\label{eq:R}
\bm{R}_{\tau} = 
\begin{bmatrix} 
	r_{1,1} & \dots & r_{1,n} \\
	\vdots & \ddots & \vdots\\
	r_{m,1} & \dots & r_{m,n} \\
\end{bmatrix}\,,
\end{align}where $m$ and $n$ denote a grid of latitude and longitude points, respectively, within the service area. Here, $r_{i,j}$ denotes the required traffic in Mbps at a location indexed by $i$ and $j$.  

In principle, the output of the CNN should encompass all potential payload configurations, which consist of feasible combinations of power and bandwidth pairs per-beam derived from the sets $(\mathcal{P}_\tau,\mathcal{W}_\tau)$. In fact, a payload configuration can be represented as $S_{\tau}=\left[ (P_{\tau}^1,{\rm W}_{\tau}^{1}), (P_{\tau}^2,{\rm W}_{\tau}^{2}), \ldots, (P_{\tau}^B,{\rm W}_{\tau}^{B}) \right]$, with each power or bandwidth variable selected from the corresponding set $(\mathcal{P}_\tau$ or $\mathcal{W}_\tau)$. However, due to the total power and bandwidth constraints defined in equation \eqref{eq:1}, numerous configurations fail to satisfy these constraints, and are therefore not viable outputs of the CNN. As we will explain, this allows us to drastically reduce the number of neurons in the output layer of the CNN, which determines the optimized configuration.

Once the input traffic demand matrix is given, the convolutional layers are employed to extract the intermediate features that are fed to fully connected layers. The output of the fully connected layers are finally fed to a classification, softmax, layer that assigns a score (probability) to every feasible resource configuration..




To reduce the number of neurons in the output layer,  we initially list all feasible choices of power and bandwidth variables according to the sets $\mathcal{P}$ and $\mathcal{W}$. Then, by enforcing the system power and bandwidth constraints, we eliminate all configurations that fail to comply with the maximum power and bandwidth thresholds. Finally, we discard configurations that were found to be ineffective during the generation of the training data process due to the traffic demand pattern. We refer to Section V for further details.


\subsection{CNN Training}

To train the parameters of the proposed CNN, we collect labels for a number of realization of the traffic matrix in (\ref{eq:R}) by running an exhaustive search method on problem \eqref{eq:1}. Although data generation is an intensive task, it is performed off-line. Once trained, the CNN is able to quickly determine the appropriate power and bandwidth allocation for each beam based on the demand pattern while minimizing payload resource consumption. 


Training aims at minimizing the mean squared error (MSE) between a predicted payload configuration, denoted as $\hat{S}$, with the corresponding ground-truth payload configuration  $S$, i.e., \begin{equation}\label{eq:S}
\mathcal{L}(\hat{S_{\tau}}, S_{\tau}) = \frac{1}{Z}\sum_{i=1}^{Z}(\hat{S}_{\tau,i} - S_{\tau,i})^2,
\end{equation}where $Z$ is the reduced number of payload configurations.

To address the minimization of the loss function, we employ standard stochastic gradient descent (SGD) algorithm with random shuffling at each training epoch and gradients evaluated via the backpropagation algorithm (see, e.g., \cite{simeone2022machine}). To prevent overfitting and determine the optimal number of training iterations, we employ early stopping. Early stopping involves monitoring the performance of the CNN on a validation dataset during training. If the validation loss does not improve for a certain number of epochs, training is halted to avoid overfitting.

\section{Neuromorphic Computing for SatCom RRM}\label{sec:npcomp}
ML-based algorithms have gained popularity due to their performance and flexibility, but their practical application is hindered by the substantial computational power required for training and inference. This limitation becomes particularly evident when considering scenarios such as deploying these models on board of satellites, where the use of one or several GPUs, which is common for modern ML architectures, is impractical due to power restrictions. To address this challenge, one promising approach involves developing more energy-efficient versions of standard ML algorithms, e.g., by employing quantization of a model's weights. In this work, we investigate a potentially more efficient alternative, SNNs, which draw inspiration from the low-power operation of biological brains. In the following section, we provide an introduction to SNNs, including training techniques and data encoding methods. We further overview the Loihi 2 chip developed by Intel, and discuss the hardware deployment of SNNs.

\subsection{Spiking Neural Networks}
In general, an SNN is a directed, possibly cyclic, network of spiking neurons. Each spiking neuron is a dynamic system with inputs and outputs given by sequences of \textit{spikes}, or binary $\{0, 1\}$, signals. SNNs comprise \textit{read-out}, or \textit{visible}, neurons, forming the network's outputs, as well as \textit{exogeneous} input neurons. It also consists of \textit{hidden} neurons, whose role is to facilitate the output of a desired spiking sequence~\cite{Skatchkovsky2021SpikingPatterns}, given a sequence of exogeneous inputs. 

For the purpose of this study, we consider SNNs with fully connected layered topologies, i.e., each neuron in one layer is connected to all of the neurons in the next. Denoting as $L$ the number of layers in the network, each layer $\ell\in \{1, \ldots, L\}$ consists of $N_\ell$ spiking neurons; we denote by $\bm{w}_\ell$ the $N_{\ell+1}\times N_\ell$ weight matrix between layer $\ell$ and layer $\ell+1$ and by $\bm{w}$ the vector of all parameters. Each neuron $k$ in layer $\ell+1$ receives inputs from the set $\mathcal{N}_\ell$ of neurons in layer $\ell$, i.e., we do not consider recurrent connections. The last layer comprises $Z$ output neurons, corresponding to the number of classes (reduced number of payload configurations).

We consider neurons following the standard spike response model (SRM)~\cite{gerstner02spiking}. At every time-step $t = 1,\ldots,T$, where $T$ denotes the temporal horizon of the task, each spiking neuron $k$ outputs a binary signal $s_{k,t}\in\{0,1\}$, with ``1'' representing the firing of a spike and ``0'' an idle neuron.

Following the SRM, each neuron $k$ maintains at every time step $t$ an internal analog state variable $u_{k,t}$, known as the \emph{membrane potential}. Mathematically, the membrane potential $u_{k,t}$ is defined by the sum of filtered contributions from incoming spikes and from the neuron's own past outputs, i.e.,
\begin{align}
	u_{k,t}=\sum_{j\in\mathcal{N}_{\ell}}w_{k,j,\ell}\cdot(\alpha_t * b_{j,t})+\beta_t * s_{k,t},
	\label{eq:potential}
\end{align}
where $w_{k,j,\ell}$ is the element $(k,j)$ of matrix $ \bm{w}_\ell$, which corresponds to the synaptic weight between neuron $j\in\mathcal{N}_{\ell}$ and neuron $k$ in layer $\ell+1$; $\alpha_t$ represents the synaptic response to a spike from the presynaptic neurons $j\in\mathcal{N}_{\ell}$ to a postsynaptic neuron $k$; $\beta_t$ describes the synaptic response to the spike emitted by the neuron itself; and $*$ is the convolution operator. Neuron $k$ outputs a spike at time step $t$ when its membrane potential $u_{k,t}$ passes some fixed threshold $\vartheta$, i.e.,
 \begin{align}
	s_{k,t}=\Theta(u_{k,t}-\vartheta).
	\label{eq:spike}
\end{align}
We refer to the review~\cite{Skatchkovsky2021SpikingCommunicationsb} for more details. 

\subsection{SNN Training}
\label{sec:snn-training}
The training loss over the parameter vector $\bm{w}$ is defined  using the training dataset $\mathcal{D} = \{(\bm{r}, \bm{y})\}$ composed of the encoded traffic requirement signals $\bm{r} = (\bm{r}_1, \dots, \bm{r}_T)$ and corresponding targets $\bm{y} = (\bm{y}_1, \dots, \bm{y}_T)$. How to obtain the encoded signals $\bm{r}$ and $\bm{y}$ from input $\mathbf{R}$ in (\ref{eq:R}) and targets $S$ in (\ref{eq:S}) will be discussed in the nex subsection.

We define the loss $\mathcal{L}_{\bm{r}, \bm{y}}(\bm{w})$ measured with respect to a data $(\bm{r}, \bm{y}) \in \mathcal{D}$ as the error between the reference signals $\bm{y}$ and the output spiking signals produced by the SNN with parameters $\bm{w}$, given the input $\bm{r}$. 
Accordingly, the loss is written as a sum over time instants $t\in\{1,\ldots,T\}$ and over the $Z$ read-out neurons as 
\begin{align} \label{eq:loss-decomp}
    \mathcal{L}_{\bm{r}, \bm{y}}(\bm{w}) &= \sum_{t=1}^T \mathcal{L}_{\bm{r}^t, \bm{y}_t}(\bm{w}) \nonumber\\
    &= \sum_{t=1}^T \sum_{k=1}^{Z} \mathscr{L}\big(y_{k,t}, f_{k}(\bm{w}, \bm{r}^t)\big),
\end{align}
where function $ \mathscr{L}\big(y_{k,t}, f_{k}(\bm{w}, \bm{r}^t)\big)$ is a local loss measure comparing the target output $y_{k,t}$ of neuron $k$ at time $t$ and the actual output $f_{k}(\bm{w}, \bm{r}^t)$ of the same neuron, given the inputs $\bm{r}^t = (\bm{r}_1, \dots, \bm{r}_t)$ up to time $t$. The notations $f_{k}(\bm{w}, \bm{r}^t)$ and $\mathcal{L}_{\bm{r}^t, \bm{y}_t}(\bm{w})$ are used as a reminder that the output of the SNN and the corresponding loss at time $t$ generally depend on the input $\bm{r}^t$ up to time $t$, and on the target output $\bm{y}_t$ at time $t$. Specifically, the notation $f_{k}(\bm{w}, \bm{r}^t)$ makes it clear that the output of neuron $k \in N_{\ell}$ is produced with the model parameters $\bm{w}$ from exogeneous input $\bm{r}^t$, consisting of all input samples up to time $t$, using the SRM \eqref{eq:potential}-\eqref{eq:spike}.

\begin{figure*}
  \centering
  \begin{subfigure}{0.49\textwidth}
    \includegraphics[width=\textwidth]{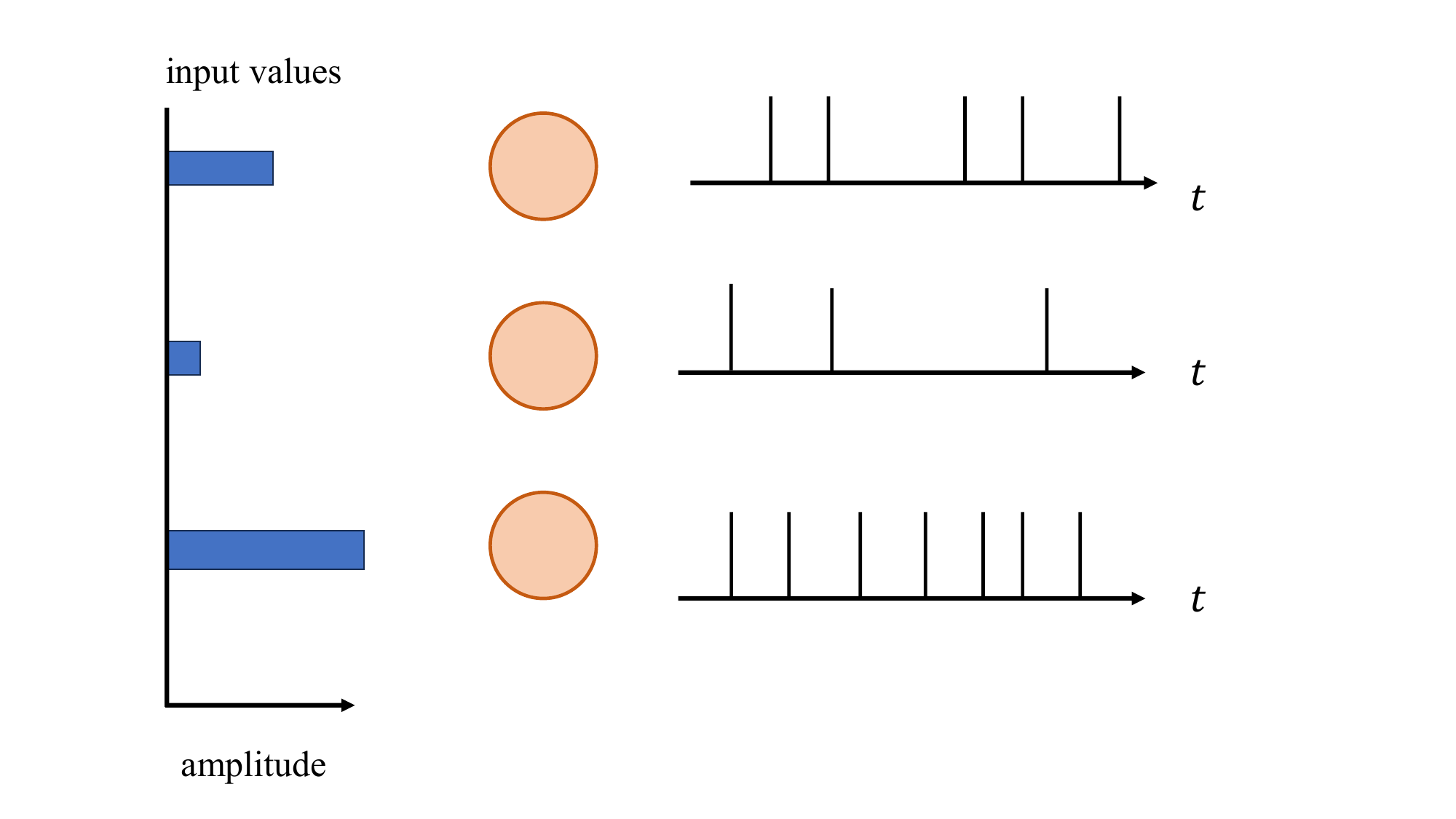}
    \label{fig:rate_coding}
  \end{subfigure}
  \hfill
  \begin{subfigure}{0.49\textwidth}
    \includegraphics[width=\textwidth]{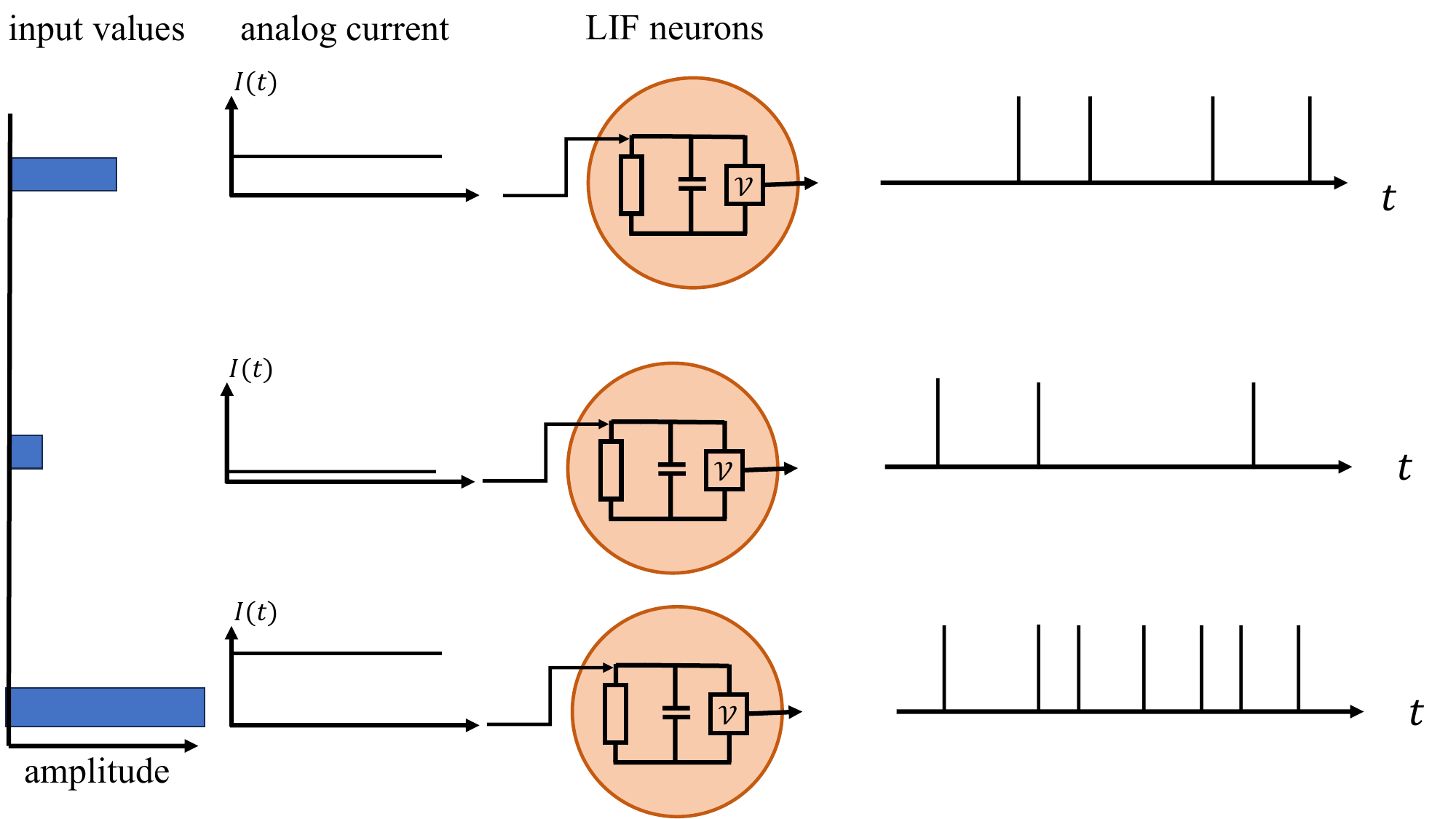}
    \label{fig:tem_coding}
  \end{subfigure}
\caption{Illustration of the encoding techniques employed for this study. Left: rate coding. Right: encoding with a time encoding machine composed of LIF neurons.} 
\label{fig:rate_encoding}

\end{figure*}

The training loss $\mathcal{L}_{\mathcal{D}}(\bm{w})$ is given as
\begin{align} \label{eq:loss}
     \mathcal{L}_{\mathcal{D}}(\bm{w}) = \frac{1}{|\mathcal{D}|} \sum_{(\bm{r}, \bm{y}) \in \mathcal{D}} \mathcal{L}_{\bm{r}, \bm{y}}(\bm{w}),
\end{align} and training is done by minimizing the loss as
\begin{align} \label{eq:erm}
     \min_{\bm{w}}~ \mathcal{L}_{\mathcal{D}}(\bm{w}).
\end{align}

We consider a \textit{spike-rate} loss, whereby the SNN is given a rate $\rho$ to dictate the desired rate of the correct output neuron for each example, and a spike rate $\rho_F$ to dictate the desired spike rate of the other output neurons. For a classification problem with $C$ classes, if $c$ is the correct class for the given input, the spike-rate loss is defined as
\begin{align}
    \label{eq:spike-rate}
    \mathcal{L}_{\bm{r}, \bm{y}}(\bm{w}) = 
    \frac{1}{2} \sum_{t=1}^{\rm T} \sum_{k \neq c} (f_{k}(\bm{w}, \bm{r}^t) - \rho_F) + (f_{c}(\bm{w}, \bm{r}^t) - \rho).
\end{align}
Throughout, we set $\rho_F = 0.01$.

Problem \eqref{eq:erm} cannot be directly solved using standard gradient-based methods since the spiking mechanism \eqref{eq:spike} is not differentiable in $\bm{w}$ due to the presence of the threshold function $\Theta(\cdot)$. To tackle the former problem, surrogate gradients (SG) methods replace the derivative of the threshold function $\Theta(\cdot)$ in \eqref{eq:spike} with a suitable differentiable approximation. We direct the reader to the reviews~\cite{Neftci2019SurrogateNetworks, Skatchkovsky2021SpikingCommunications} for more details on this approach. 

\subsection{Spike Encoding}
\label{sec:encoding}

Although spiking neuron models can in principle receive data in the form of an analog input current, neuromorphic processors can typically  handle data only in the form of binary inputs.
Consequently, the natural signals $(r_{i,j})$ representing the required traffic to be encoded into binary spikes for processing using the neuromorphic chip. 
Encoding into spiking signals is performed as follows and as illustrated in Fig.~\ref{fig:rate_encoding}. 

Considering the feature matrix $\bm{R} \in \mathbb{R}^{n \times m}$ in (\ref{eq:R}), we first perform max-pooling with pool size $(ds, ds)$ to obtain an $(n/ds) \times (m/ds)$ matrix. This dimensionality reduction was key to reduce the transmission time of data to the neuromorphic chip.
The resulting matrix is flattened into $(n/ds)(m/ds) \times 1$ column vector $[r_{1, 1} \dots r_{n/ds,m/ds}]^{\rm T}$. We finally perform the encoding of this column vector into a collection of spiking signals $\bm{r} \in \{0, 1\}^{(n/ds)(m/ds) \times T}$, with $T$ being the number of encoding time-steps. Although a number of encoding techniques are used in the literature, we propose to compare in this study results obtained using rate encoding~\cite{Skatchkovsky2021SpikingPatterns} and a time encoding machine (TEM). We detail both techniques below. 


\textit{(i) Rate encoding:} As seen in Fig.~\ref{fig:rate_encoding}, each input value, as shown using the horizontal blue bars on the left, is encoded in the spike rate of the corresponding encoding neuron: a larger input generates a large number of spikes within a fixed encoding window time. 

\textit{(ii) Time encoding machine:} A TEM is a system that receives as input a (bounded) natural signal $x(t)$, and outputs binary spikes. We consider a TEM model based on leaky integrate-and-fire (LIF) neurons, whereby a spike is emitted when the voltage of the neuron crosses a pre-defined threshold. More specifically, the TEM operates using the following set of recursive equations
\begin{align}
        \label{eq:tem-operation}
        u_t &= (1 - \alpha_u) \cdot u_{t-1} + x_t  \nonumber\\
        v_t &= (1 - \alpha_v) \cdot v_{t-1} + u_t \nonumber \\
        s_t &= \Theta(v_t-\vartheta)\nonumber \\
        v_t &= v_t \cdot (1-s_t),
\end{align}
where $u_t$ and $v_t$ denote the current and voltage of the neuron at time instant $t$;  $x_t := x(t)$; $1 > \alpha_u > 0$ and $1 > \alpha_v > 0$; $\vartheta$ is the threshold of the neuron; $s_t \in \{0,1\}$ is the spike output at instant $t$; and $\Theta(\cdot)$ is the Heaviside step function. In practice, a single signal $x$ can be encoded through $N$ TEMs with varying decays $\alpha_u$ and $\alpha_t$. 

\section{Chipset Implementation}

In this section, we present the implementation of the CNN model and the SNN model in hardware chipsets, namely Xilinx Versal VCK5000 and Intel Loihi2, respectively.

\begin{table}[t]
\caption{Simulation parameters for generating training data}\label{table:parameters}
\begin{tabular}{|l|l|}
\hline
Frequency          & 19 GHz                                                                                                                         \\ \hline
Satellite position & 13 E                                                                                                                           \\ \hline
Satellite altitude & 35786 km
                                                \\ \hline
$\theta_{\rm 3\,dB}$         & 1 deg.                                                                                                                         \\ \hline
Number of beams    & 8                                                                                                                              \\ \hline
Beam centres       & \begin{tabular}[c]{@{}l@{}}lat =[39.3 42 44.7 47.4 51 53.7 56.4 39.5] \\ long =[-5.3 0 5.3 10.6 -0.5 6 12.3 14.4]\end{tabular} \\ \hline
$G/T_{\rm RX}$             & 17 dB/K                                                                                                                        \\ \hline
$P_t^b$             & 10, 12, or 14 dBW                                                                                                                      \\ \hline
${\rm W}_t^b$             & 250 or 500 MHz                                                                                                                     \\ \hline
\end{tabular}
\end{table}

\begin{table*}[ht]
\centering
\caption{Possible resource allocation configurations in a beam}\label{tab:capcity_beam}
\begin{tabular}{|l|l|l|l|l|l|}
\hline
${\rm W}_{\tau}^b$ [MHz] & $P_{\tau}^b$ [dBW] & ${\rm EIRP}_{\max}^b$ [dBW] & ${\rm \gamma}_{\tau}^b$ [dB] & ${\rm \kappa}_{\tau}^b$ [bps/Hz] & $C_{\tau}^b$ [Mbps] \\ \hline
250              & 10            & 54.94           & 6.4615    & 1.8865        & 471.6312         \\ \hline
500              & 10            & 54.94            & 3.4817    & 1.3350        & 667.4827        \\ \hline
250              & 12            & 56.94            & 8.4261    & 2.2502        & 562.5421        \\ \hline
500              & 12            & 56.94            & 5.4638    & 1.7019        & 850.9288        \\ \hline
250              & 14            & 58.94            & 10.3705   & 2.6101        & 652.5215       \\ \hline
500              & 14            & 58.94            & 7.4357    & 2.0668        & 1033.400          \\ \hline
\end{tabular}
\end{table*}

First, let us define the simulation parameters. The focus of this study is on downlink analysis for the forward link. The parameters used during the simulations are listed in Table~\ref{table:parameters}. We consider two different values of bandwidth, $\mathcal{W}_\tau = \{250, 500\}$ MHz, and three values of power, $\mathcal{P}_\tau =\{10, 12, 14\}$ dBW. This results in six possible values of capacity that can be offered in each beam as listed in  Table~\ref{tab:capcity_beam}.

We focus on eight beams and six configurations per beam, and hence the number of possible payload configurations is more than 40,000 options. However, after setting $P_{\max,T}= 115$ W, all configurations that do not meet the power constraint are eliminated, resulting in the reduction of the number of feasible configurations to less than 1\%. Finally, many configurations will be discarded again due to the traffic pattern as explained in Section III, resulting in $Z=6$ distinct possible classes (number of configurations that are actually deployed on the satellite). 

30,000 samples were generated, The data was divided into two sets, with 80\% used for training and 20\% used for validation. 



\subsection{CNN Model}

The CNN model is implemented on the VCK5000 AI accelerator, a high-performance platform based on the Xilinx 7nm Versal ACAP architecture. It utilizes matrices as input and applies convolutional layers for feature extraction, followed by fully connected layers for classification \cite{Ortiz-Gomez2021ConvolutionalSystems}. The CNN architecture comprises Conv2D, Maxpooling2D, and Dense layers.

The specific CNN architecture employed consists of Conv2D layers with 8 filters and a kernel size of (3,3), followed by Maxpooling2D layers with a pool size of (2,2). This is followed by additional Conv2D and Maxpooling2D layers with 4 filters and a kernel size of (3,3) and (2,2), respectively. The subsequent layers include a Flatten layer, Dense layers with 512 and 256 units and ReLU activation, and a final Dense layer with 6 units and softmax activation, representing the different payload configurations. The trainable parameters sum up to a total of 3,192,058.

To implement the CNN model, we consider using the VCK5000 Versal development card, designed to provide high-throughput AI inference and signal processing compute performance. It supports popular ML frameworks such as TensorFlow, PyTorch, and Caffe, using Python or C++ APIs. The Vitis AI framework facilitates the deployment of TensorFlow/PyTorch trained models on the VCK5000 for inference. In Table~\ref{tab:CardFeatures} we summarize the main VCK5000 card features.
\begin{table*}[ht]
 \centering
 \setlength{\tabcolsep}{3pt}
 \caption{Summary of VCK5000 card features}\label{tab:CardFeatures}
    \begin{tabular}{|p{200pt}|p{250pt}|}
        \hline
        \textbf{Feature} &  \textbf{Description} \\
        \hline
            ACAP Device & XCVC1902-2MSEVSVD1760 \\
            \hline
            Configuration Options & 2 GB OSPI memory MT35XU01GBBA2G12-0SIT\\
            \hline
            Memory & 16 GB - four 4 GB modules, part MT40A512M16TB-062E:J \\ 
            \hline
            DDR Maximum Data Rate & 3200 MT/s \\
            \hline
            DDR Memory Bandwidth & 102.4GBs \\
            \hline
            JTAG and UART Debug Interface  & JTAG and UART access through USB and Maintenance connector \\
            \hline
            Edge Connector Interface & PCIe Gen. 3 x16 / 2x Gen. 4 x 8 / Gen. 4 x 8 /CCIX \\
            \hline
            Network Interface & 2x QSFP28 \\
            \hline
            ACAP VCCINT VRM  & 6 phases. PMBUS control for VOUT, temperature, and current \\
            \hline
            Satellite Controller (SC) & MSP432P4111IPZR\\
            \hline
            Temperature Monitoring & ACAP (internal monitoring), QSFP, inlet and outlet on-board
thermal sensors\\
            \hline
            Power Monitoring & PCIe 12V, AUX0, AUX1, and VRM VCCINT $\&$ 1.2V \\
            \hline
            Status LEDs & Activity LED\\
            \hline
            Form Factor & Passive configuration: 3/4-length, full height, double slot, x16 PCIe
form factor.
Active configuration: full length, full height, double slot, x16 PCIe
form factor\\
        \hline
            Motherboard & PCIe 3.0-compatible with one dual-width x16 slot. \\
            \hline
            System Power Supply & Minimum 225W available via PCIe slot connection and 6-pin and 8-pin PCIe
auxiliary power cables\\
            \hline
            Operating System & Linux, Windows\\
            \hline
            Thermal Options & Active or Passive\\
            \hline
    \end{tabular}
\end{table*}


\subsection{Neuromorphic Model for RMM}
\label{sec:payload-snn}

\begin{figure*}
  \centering
     \includegraphics[width=0.85\textwidth]{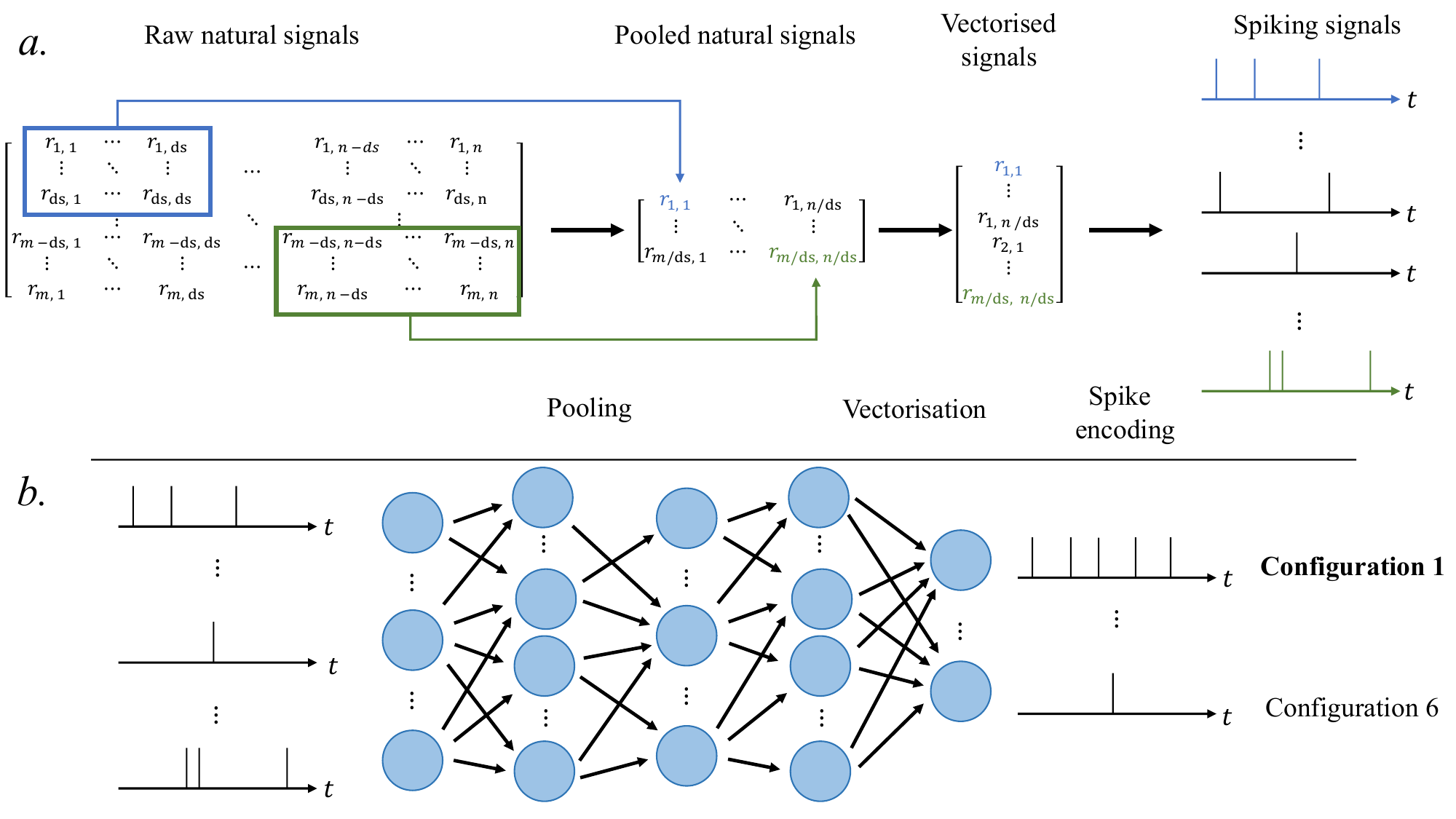}
  
  \caption{Proposed neuromorphic solution. (a) Encoding of natural signals into spikes. (b) Spiking neural networks with a layered architecture comprising three hidden layers. The prediction is given by rate decoding, that is, by selecting the index of the readout neuron producing the most spikes.}
  \label{fig:SNN_RRM}
\end{figure*}


As seen in Fig.~\ref{fig:SNN_RRM}, we consider a layered SNN with $L=4$ layers, where the hidden layers comprise $512, 256$ and $512$ neurons respectively, and $Z=6$. We train the system via the SG-based method SLAYER~\cite{shrestha18slayer}. SGD is carried out using the Adam optimizer. Models are trained using Intel's Lava library~\cite{loihi2} with Loihi bit-accurate precision, on a single A100 GPU. On-chip training was not yet available on Intel's Loihi 2 at the time of writing. Decisions are obtained via rate decoding, i.e., by selecting the output neuron with the largest spiking rate.

\begin{figure*}
\centering
\includegraphics[width=\linewidth]{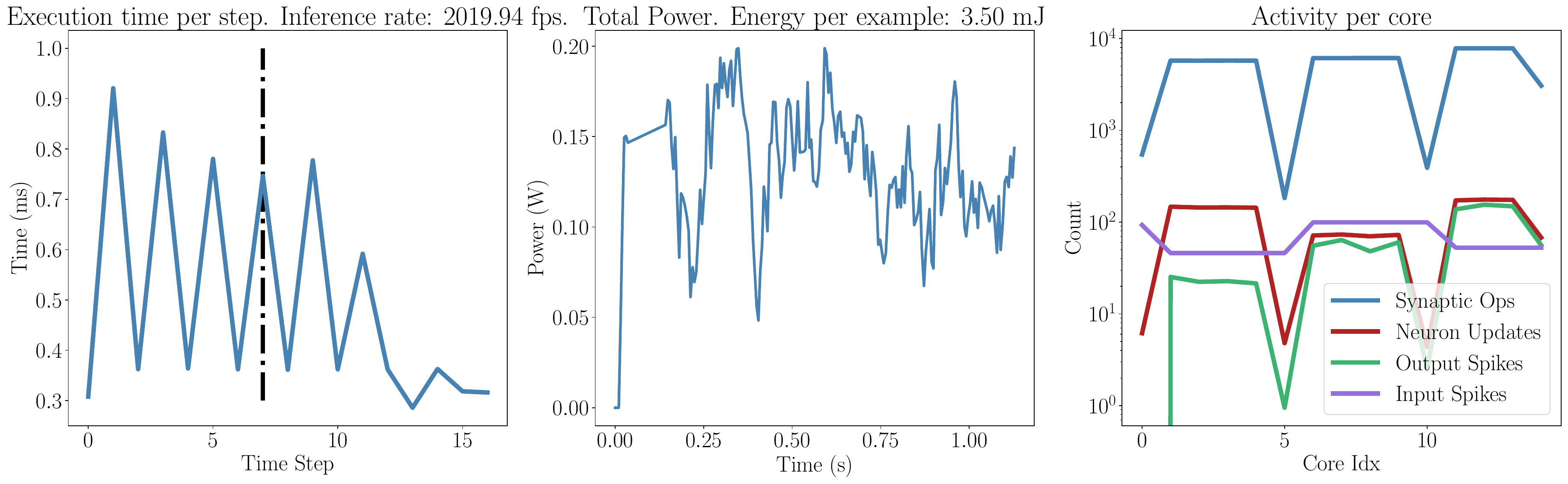}
\caption{Summary of the metrics measured on Loihi 2. Left: Execution time per algorithmic time-step. Middle: Evolution of the total power consumption. Right: Activity per core.}\label{fig:payload-measurements}
\end{figure*}

Following the approach proposed in reference~\cite{Ortiz-Gomez2020SupervisedSystems}, training is completed using a dataset $\mathcal{D}$, composed of measurements of the required capacity in each geographical zone. Each example $\bm{R} \in \mathbb{R}^{m \times n}$ in dataset $\mathcal{D}$ consists of $n \times m$ resource requirements, in Mbps, for each geographical position, as detailed in Section~\ref{sec:system}. We preprocess each example independently as follows. First, we set all the outlier values over a given percentile $p$ to the value $r_p$, which is the value such that $p \%$ of the entries in $\bm{R}$ are smaller than $r_p$. We then normalize the examples to the range $[0, 1]$, and perform max-pooling with stride $ds$ to reduce the input size to $(n/ds) \times (m/ds)$ before encoding into binary spiking signals, as described in Section~IV.\ref{sec:encoding}.

We perform inference using SNNs as described in the previous sections on Intel's Loihi 2 chips~\cite{loihi2}. Loihi 2 is a research neuromorphic chip that uses asynchronous spiking neurons to implement fine-grained, event-driven, adaptive, self-modifying, parallel computations. Loihi's first iteration was fabricated on Intel's 14 nm process and houses 128 clusters of 1,024 artificial neurons each, for a total of 131,072 simulated neurons, which is about 130 million synapses, which is still far below the 800 trillion synapses in the human brain. As members of the Intel Neuromorphic Research Community (INRC), we were given access to Loihi 1 under the Kapoho Bay form factor (see Table~\ref{tab:Loihi}), as well as the second iteration of the chip via Intel's cloud services. Experimental results were obtained on Loihi 2.

\begin{table}[t]
\setlength{\tabcolsep}{3pt}
    \centering
    \caption{Summary of Kapoho Bay USB Flash Drive incorporating 2 Loihi chips}
    \label{tab:Loihi}
    \begin{tabular}{|l|l|} 
    \hline 
    \textbf{Feature} & \textbf{Description} \\ 
    \hline 
    Form factor & USB stick \\
    \hline 
    Host interface & USB 3.0 \\ 
    \hline 
    DVS interface & For neuromorphic sensors such as a camera \\
    \hline 
    Loihi chips & 2 \\ 
    \hline 
    Neuromorphic cores & 256 \\
    \hline 
    Artificial neurons & 262,144 \\
    \hline 
    Synapses & 260 million \\
    \hline 
    Process & 14 nm \\
    \hline 
    Transistors & 4.14 billion \\
    \hline 
    \end{tabular}
\end{table}

The Lava software library gives access to a number of metrics of interest, of which we show a summary in Fig.~\ref{fig:payload-measurements}. We can hence measure the evolution of the execution time per algorithmic step, and the total power consumed by them, as well as the activity over the various cores on the chip. Activity measurements comprise the number of synaptic operations, neuron updates, and output and input spikes per core. As can be seen, the execution time per step decreases after the first few steps. In our experiments, we exploit this by initializing the network by sending all-zeros inputs up to $t=8$ time-steps, and start recording predictions from that point onwards, with the black dotted line marking the beginning of the recording. During inference for the task at hand, the power consumption is seen to vary around $0.12$~W, which comprises the power expenditure for input and output of spikes to the chip. Finally, the activity of the network is balanced over several cores, which is directly optimized by the Lava library.

\subsection{Evaluation Metrics}
We evaluate the performance of the proposed algorithm in terms of \textit{average capacity gap}. This is a measure of the capacity gap between the predicted configuration, and the resource requirements. Formally, it is defined as
\begin{align}
    G = \frac{1}{B|\mathcal{D}|} \sum_{\bm{R} \in \mathcal{D}}\sum_{b = 1}^{B} \big| C_{\bm{R}}^{b} - Y_{\bm{R}}^b \big|,
\end{align}
where we have defined $C_{\bm{R}}^{b}$, the required capacity for beam $b$ in example $\bm{R}$, and the corresponding prediction $Y_{\bm{R}}^b$.

Further metrics can be employed when comparing a neuromorphic to a conventional approach. These metrics include accuracy, but also latency and energy consumption, as we detail now.
\begin{itemize}
    \item \emph{Accuracy}: Accuracy measures the ability of the RRM algorithm to choose the most appropriate configuration given the current traffic requirements. It is obtained by comparing the configuration predicted by the model to the ground truth. Comparing the accuracy achieved by both approaches allows to determine which approach performs better in effectively utilizing the available resources.
    \item \emph{Latency}: Latency is a measure of the computational efficiency and responsiveness of the algorithm under study, and a key criterion in many SatCom applications. In the case of RRM, it measures how quickly the proposed algorithm can measure a change in the traffic requirement conditions, and propose an alternative configuration.
    \item \emph{Energy Consumption}: Energy consumption is a critical metric, especially in satellite systems, wherein power resources are often limited. Evaluating the energy consumed by RRM approaches allows us to assess their suitability for deployment on board of satellites. The energy consumption of the neuromorphic solution includes all the energy spent by the chip during inference, including I/O interfaces. For the conventional approach, energy consumption relates to the power consumed by the VCK chip during inference. Comparing the energy consumed by both approaches allows to inform choices pertaining to the deployment of algorithms on board of satellite systems.
\end{itemize}

\section{Experimental Results}
In this section, we present the most outstanding numerical results.

\subsection{Benchmarking}

\begin{figure*}[ht]
\centering
\includegraphics[width=\linewidth]{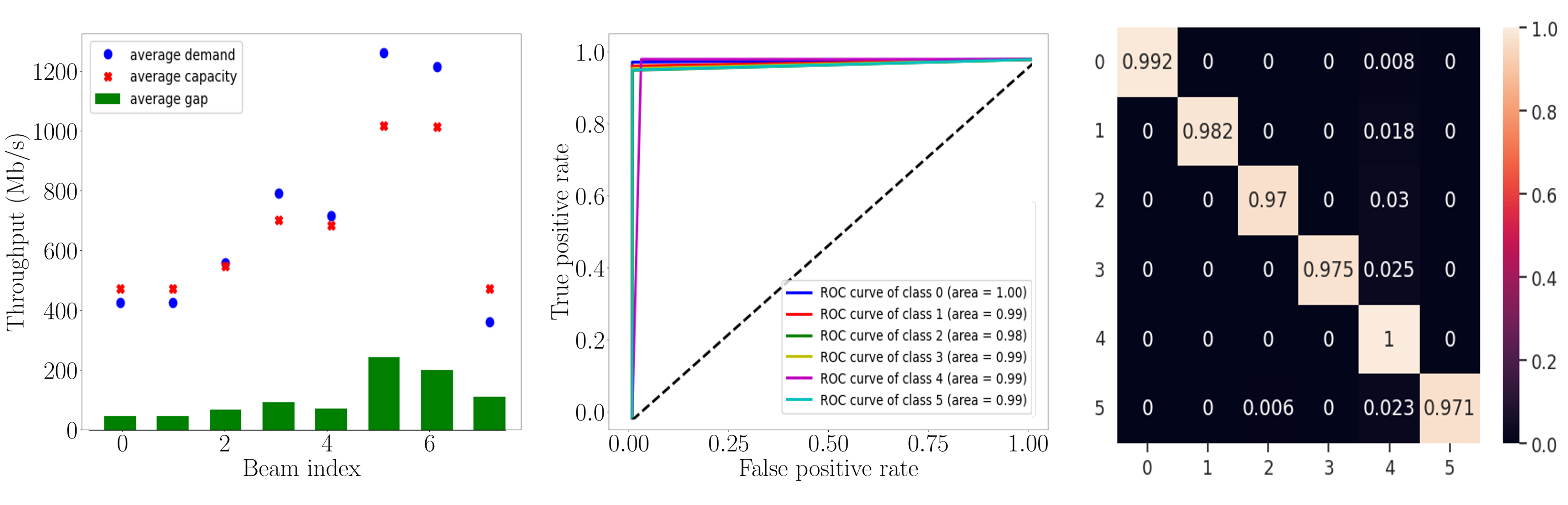}
\caption{Results obtained for a CNN model with max-pooling stride $ds = 3$. Left: Capacity gap per beam. Middle: ROC curves. Right: Confusion matrix.}\label{fig:payload-resultsCNN}
\end{figure*}

The CNN model was trained for 25 epochs with a batch size of 128. The training for the conventional models was carried out using the HPC facilities of the University of Luxembourg~\cite{VBCG_HPCS14}
(see \href{http://hpc.uni.lu}{hpc.uni.lu}). The training configuration used the SGD optimizer with a learning rate of 0.01, momentum of 0.9, and Nesterov acceleration. The loss function used was the categorical cross-entropy and the model's accuracy was used as the metric. 

We determined experimentally that the set of hyperparameters providing the maximal accuracy. The final accuracy of the model on the training data corresponded to max-pooling stride $ds = 3$ was 98.93\% and the validation accuracy was 98.87\%. In Fig.~\ref{fig:payload-resultsCNN}, we show the capacity gap, receiver operating characteristic (ROC) curves and confusion matrix for the predictions given by this model. The capacity gaps stem from the power constraint of the system, which does not allow to serve all the users. Concerning the system performance, Fig.~\ref{fig:payload-results} shows the average capacity gap in the validation data in which we observe that in all cases the capacity follows the traffic demand and in the cases where the gap is larger, it is due to the lower and upper limit on the satellite resources. In this regard, the confusion matrix shows that for all configurations we obtain a ratio higher than 96\%, reaching a ratio of 100\% in class 4, which is the class that has the highest probability of occurrence. 

The ROC plots the true positive rate (TPR) against the false positive rate (FPR) at different classification thresholds. It provides information about the trade-off between TPR and FPR, and it is useful for selecting the optimal classification threshold. The area under the curve (AUC) is the area under the ROC curve. AUC ranges between 0 and 1, where a higher value indicates a better model. Accordingly, we can observe that the CNN results for the flexible payload maintain an area close to 1 for all classes. 

As for the F1-score, it measures the model's balance between precision and recall. A high F1-score indicates a good balance between the two. Table~\ref{tab:F1_CNN_FP} shows the obtained F1-scores. The results indicate that the CNN architecture and the training configuration used were able to effectively manage the power and bandwidth of a multibeam satellite as a function of traffic demand.

\begin{table} [ht]
\caption{F1-score for test data}\label{tab:F1_CNN_FP}
  \centering
  \begin{tabular}{ |c|c| }
 \hline
 \textbf{Classes} & \textbf{F1 Score}  \\
 \hline
  \hline
 Configuration 0 & 0.996   \\
 \hline
 Configuration 1 & 0.990  \\
 \hline
 Configuration 2 & 0.972  \\
 \hline
 Configuration 3 & 0.987  \\
 \hline
 Configuration 4 & 0.991  \\
 \hline
 Configuration 5 & 0.985  \\
 \hline
 \end{tabular}
\end{table}

Finally, we vary the size of the traffic matrix, $\bm{R}_{\tau}$, by considering several max-pooling strides $ds$. As seen in Table \ref{tab:acc_maxpooling}, by increasing $ds$ (i.e., reducing the matrix size), the execution time can be reduced. However, this comes at the cost of accuracy. Hence, we use $ds=3$ network for the purpose of comparison with the neuromorphic model in the following sections.

\begin{table} [ht]
\caption{Accuracy and time per example for different size of the traffic matrix using maxpooling}\label{tab:acc_maxpooling}
  \centering
  \begin{tabular}{ |c|c|c|}
 \hline
 \textbf{Max-pooling strides ds} & \textbf{Accuracy} & \textbf{time per example (ms)}  \\
 \hline
  \hline
 No max-pooling & 0.988 & 126 \\
 \hline
 $ds=3$ & 0.988 & 114 \\
 \hline
 $ds=28$ & 0.962 & 68\\
 \hline
 $ds=30$ & 0.957 & 67\\
 \hline
 $ds=32$ & 0.951 & 67\\
 \hline
 $ds=34$ & 0.945 & 64\\
  \hline
 $ds=36$ & 0.931 & 64\\
 \hline
 \end{tabular}
\end{table}

\subsection{Neuromorphic Model}
\label{sec:payload-snn-results}
\begin{figure*}[ht]
\centering
\includegraphics[width=\linewidth]{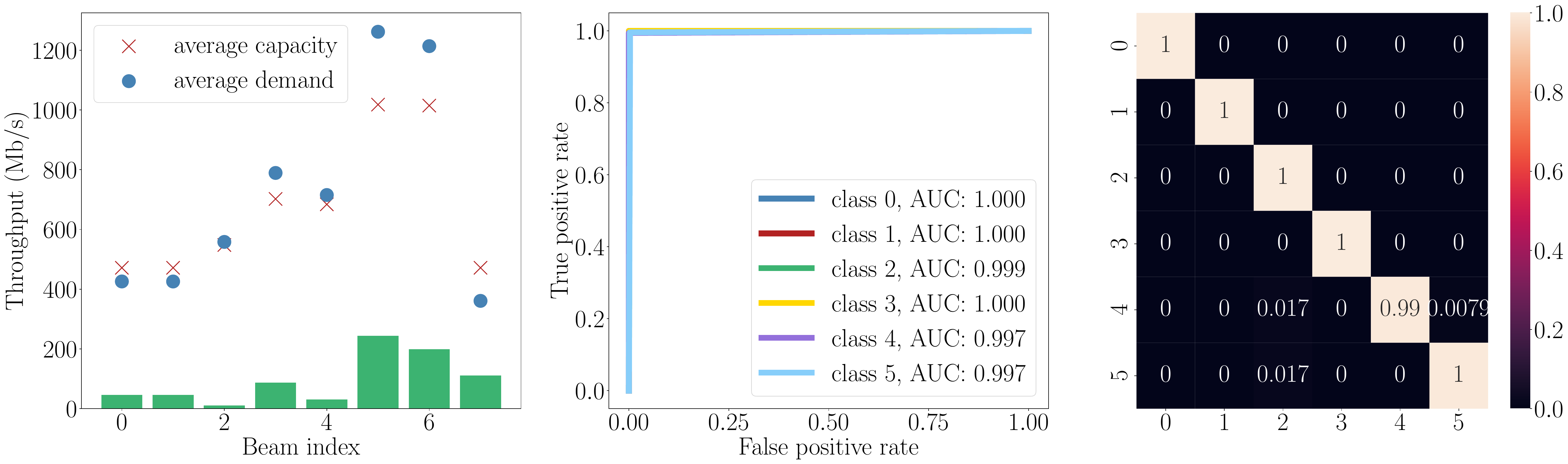}
\caption{Results obtained for an SNN model with max-pooling stride $ds=32$; number of encoding time-steps $T=8$; target spike rate $r = 0.5$, with TEM encoding. Left: Capacity gap per beam. Middle: ROC curves. Right: Confusion matrix.}\label{fig:payload-results}
\end{figure*}

We determined experimentally that the set of hyperparameters providing the maximal accuracy whilst minimizing energy expenditure and execution time corresponded to max-pooling stride $ds=32$; number of encoding time-steps $T=8$; and target spike rate $\rho = 0.5$, when using TEM encoding. This set of hyperparameters provides a reference model, which reaches $99.6\%$ of test accuracy. In Fig.~\ref{fig:payload-results}, we show the capacity gap, ROC curves and confusion matrix for the predictions given by this model. As in the results observed with the CNN model, the capacity gaps stem from the power constraint of the system, which does not allow to serve all the users. The predictions made by the SNN provide better AUC than the CNN for all classes, only predicting the wrong classes for labels $4$ and $5$, contrary to the conventional solution, which seldom reach perfect accuracy for a single class. Overall, the SNN is seen to obtain better accuracy than the CNN.

\begin{figure*}
\centering
\includegraphics[width=\linewidth]{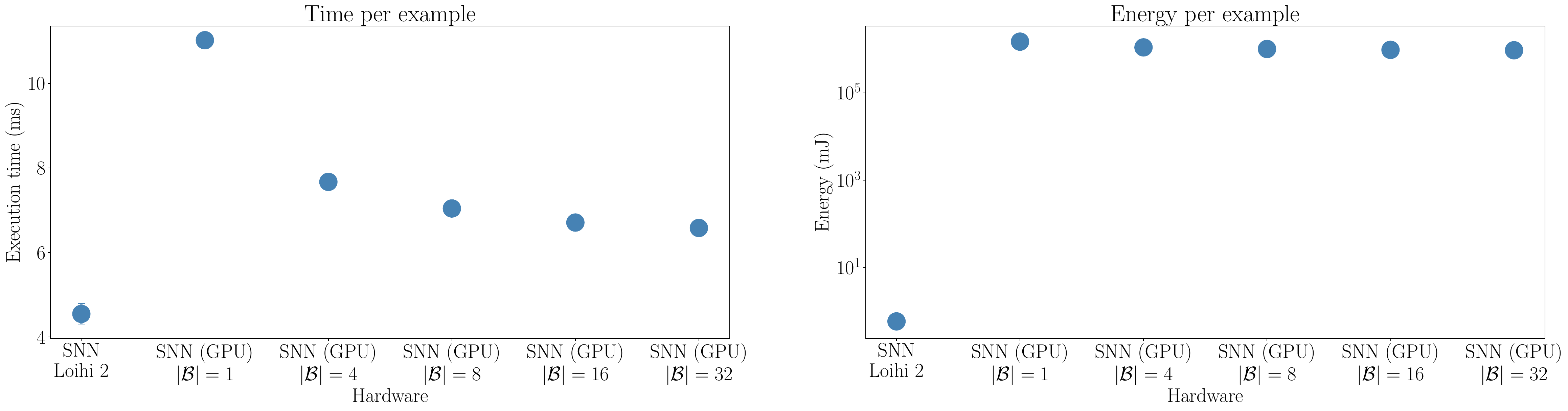}
\caption{Comparison between execution of an SNN on Loihi 2 and simulation on an NVIDIA A100 GPU with various batch sizes. Left: Average execution time per example. Right: Energy expenditure.}\label{fig:payload-gpu}
\end{figure*}

 In Fig.~\ref{fig:payload-gpu}, we compare results in terms of average execution time and energy expenditure per example between the neuromorphic processor and Intel's \textit{lava-dl} simulator running on an A100 GPU. Measurements on the GPU are carried using the PyJoules library~\cite{pyjoules}. It can be seen that the simulator running on GPU is slower than the execution on Loihi 2, but also six orders of magnitude more costly in terms of energy. It is to be noted that execution on GPU also allows one to perform inference over a minibatch $\mathcal{B}$ of size larger than one, which is not possible on the neuromorphic hardware.

\begin{figure*}[ht]
\centering
\includegraphics[width=\linewidth]{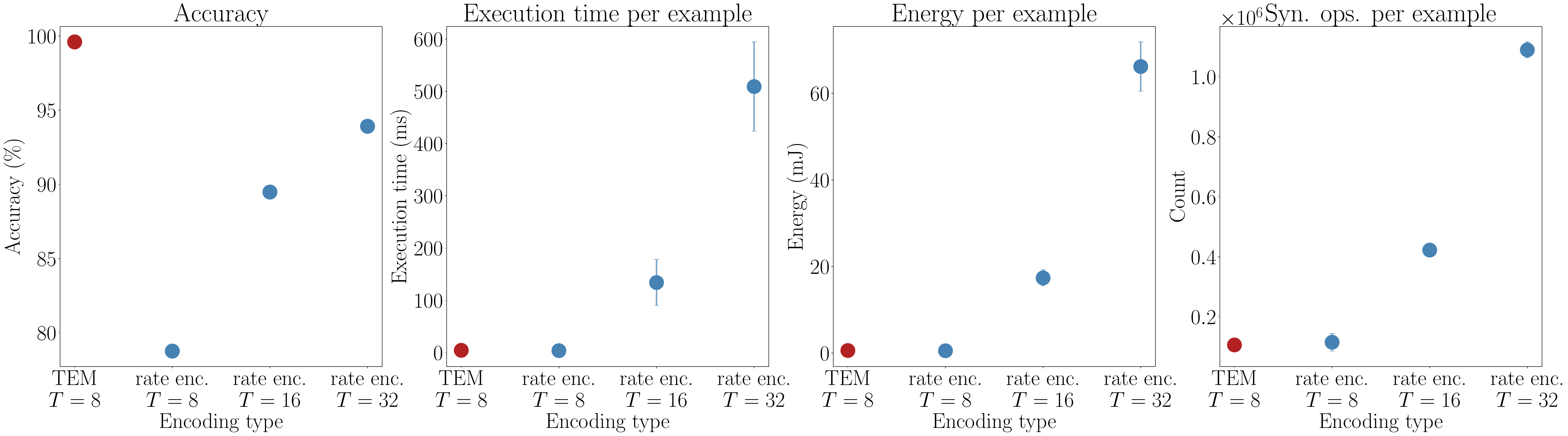}
\caption{Comparison between various types of encoding techniques and number of encoding time-steps. From left to right are shown: Test accuracy, average execution time per example, average energy expenditure per example, average number of synaptic operations per example.}\label{fig:payload-encoding}
\end{figure*}

In the following, we extensively study how the different hyperparameters of the system affect the metrics of interest -- that is, accuracy, execution time, energy and the number of synaptic operations. In all the following figures, the red data point corresponds to the model with the hyperparameters of reference described above.

We start by comparing the two encoding techniques described in Section~\ref{sec:npcomp}.\ref{sec:encoding}, i.e., rate encoding and time encoding machines. As can be seen in Fig.~\ref{fig:payload-encoding}, the TEM generally provides much higher accuracy, at the cost of a slightly larger execution time and energy expenditure compared with rate encoding with $T=8$. By increasing the number of encoding time-steps, one can improve the accuracy of the model for rate encoding, although this proportionally increases execution time, energy expenditure and number of synaptic operations. Even with a $4\times$ longer encoding of $T=32$, rate encoding only reaches $85.6\%$ test accuracy, which is $14$ points below the results obtained by encoding the signal with a TEM.

\begin{figure*}[ht]
\centering
\includegraphics[width=\linewidth]{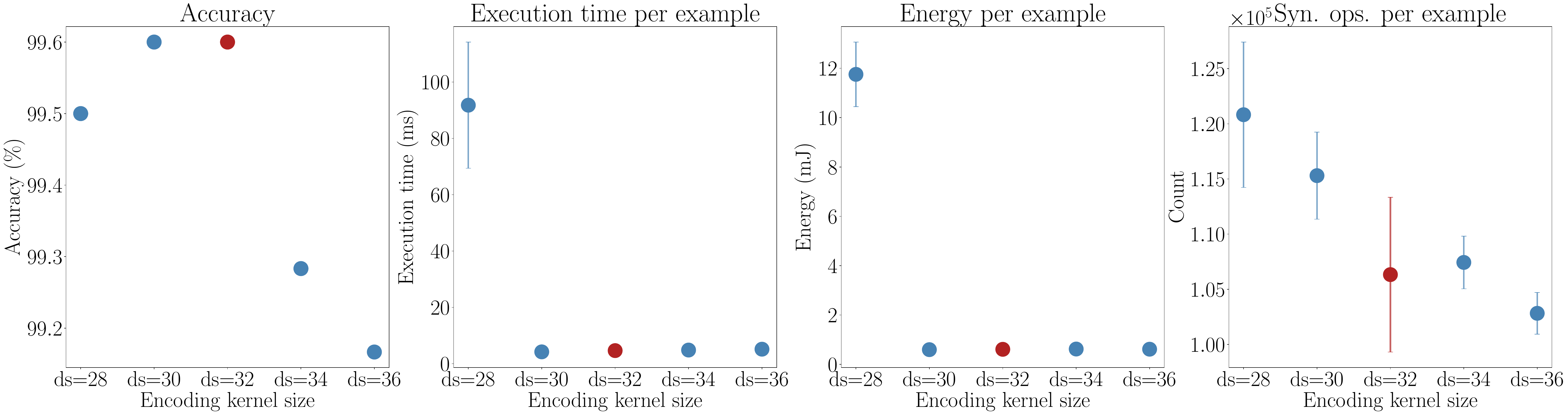}
\caption{Comparison between various encoding max-pooling strides $ds$. From left to right are shown: Test accuracy, average execution time per example, average energy expenditure per example, and average number of synaptic operations per example.}\label{fig:payload-ds}
\end{figure*}

Next, we vary the size of the exogenous inputs by considering several max-pooling strides $ds$. As seen in Fig.~\ref{fig:payload-ds}, by increasing the value of $ds$ (that is, by reducing the exogenous signal size), one can reduce the execution time, energy and synaptic operations per example. However, this comes at the cost of accuracy, By reducing $ds$ from $ds=28$ to $ds=36$, which results in decreasing the size of the inputs from $(n/22)\times(m/22) = 299$ to $(n/30)\times(m/30) = 180$, the test accuracy only diminishes by less than $1\%$.

\begin{figure*}[ht]
\centering
\includegraphics[width=\linewidth]{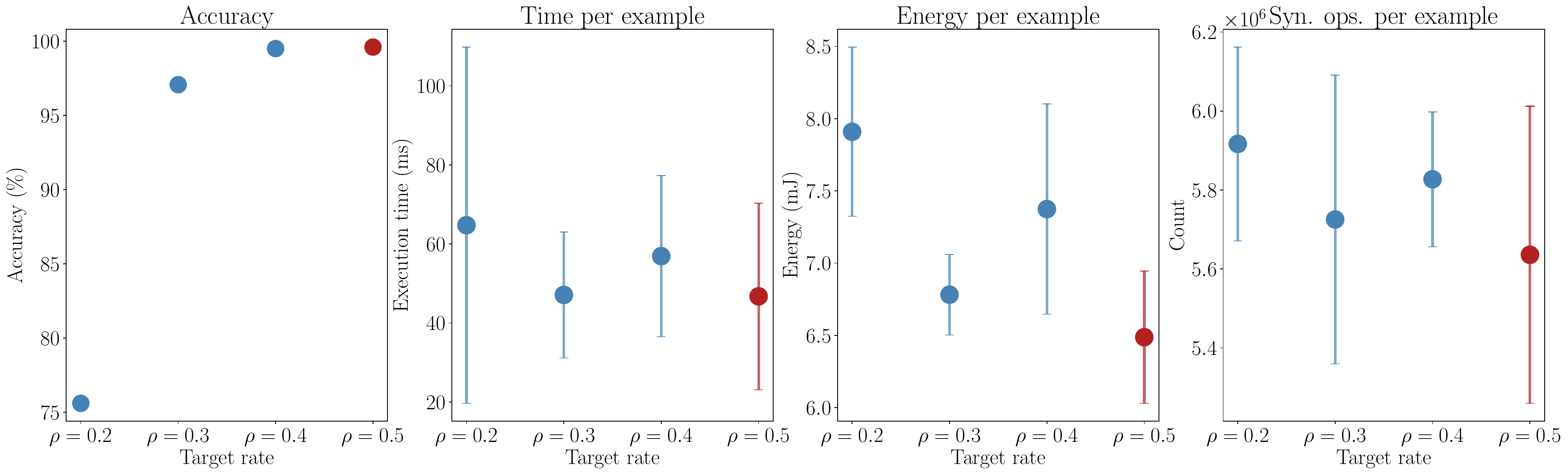}
\caption{Comparison between various target rates $\rho$. From left to right are shown: Test accuracy, average execution time per example, average energy expenditure per example, and average number of synaptic operations per example.}\label{fig:payload-rate}
\end{figure*}

In Fig.~\ref{fig:payload-rate}, we demonstrate how using smaller target rates $\rho$ during training affects the metrics of interest. As one can see, using a rate smaller than $\rho = 0.5$ generally increases the processing time, energy per example, and the number of synaptic operations. This can be explained by the fact that forcing a smaller output rate actually pushes the hidden neurons to spike more to encode the signals of interest. Considering the fact that these are more numerous than the output neurons, this results in a generally larger number of synaptic operations, which also increases the processing time and energy expenditure. We determined experimentally that $\rho = 0.5$ provided the best trade-off between sparsity of the read-out neurons, sparse activation of the hidden neurons, and test accuracy. 

\begin{figure*}[ht]
\centering
\includegraphics[width=\linewidth]{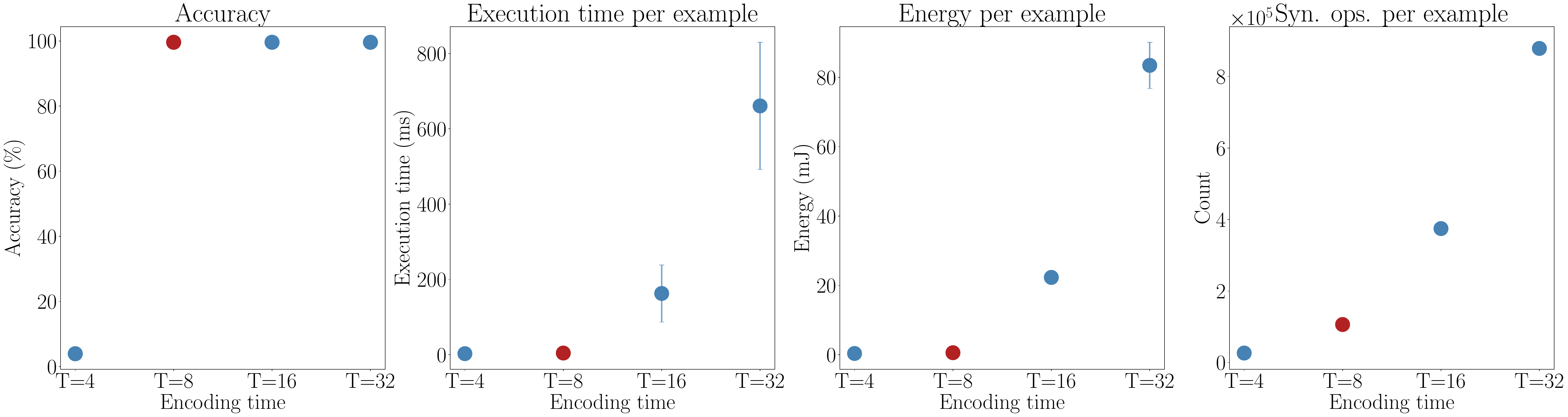}
\caption{Comparison between the number of encoding time-steps $T$. From left to right are shown: Test accuracy, average execution time per example, average energy expenditure per example, and average number of synaptic operations per example.}\label{fig:payload-time}
\end{figure*}

We now examine how the number of encoding time-steps $T$ affects the model when encoding the input signals with a TEM. As can be seen, decreasing $T$ below the reference value $T=8$ provides worse-than-chance accuracy. This is due to the fact that with $T=4$, the encoder does not produce enough spikes for the model to carry out the classification task. Increasing the number of time-steps does not provide an improvement in terms of accuracy. 

\begin{figure*}[ht]
\centering
\includegraphics[width=\linewidth]{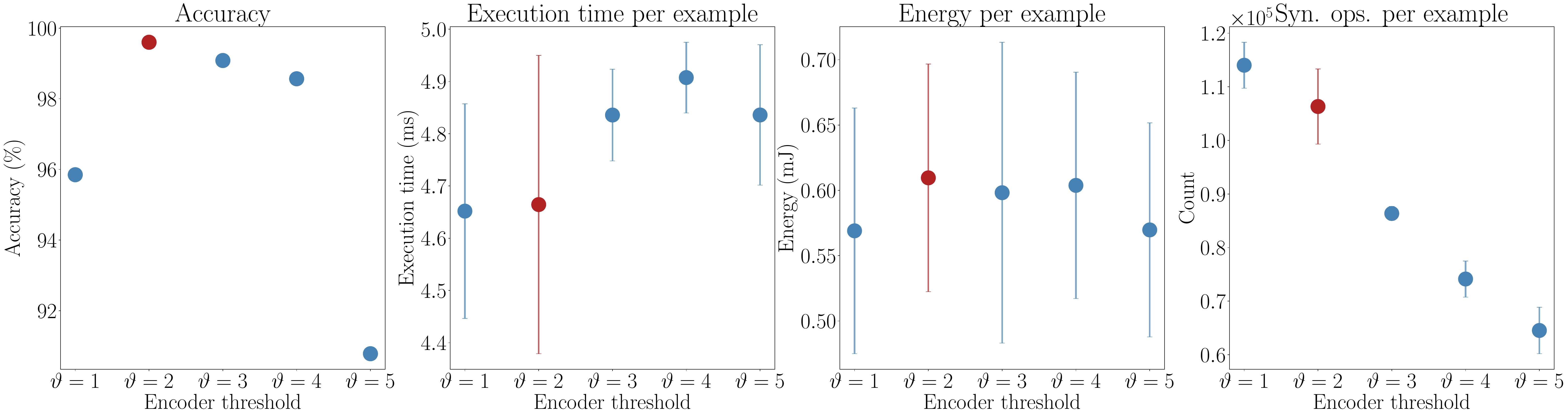}
\caption{Comparison between the number of TEM threshold $\vartheta$. From left to right are shown: Test accuracy, average execution time per example, average energy expenditure per example, and average number of synaptic operations per example.}\label{fig:payload-thr}
\end{figure*}

Lastly, varying the threshold $\vartheta$ of the TEM in \eqref{eq:tem-operation} allows to control the sparsity of the input signal. We observe that increasing $\vartheta$ generally decreases the accuracy of the system, although the decrease is almost negligible for $\vartheta < 5$. As one can expect, sparser rates in the exogenous inputs in turn cause the network to spike less, which results in a smaller energy footprint. However, this does not bring benefits in terms of execution time per example. 

Overall, we explored how a variety of hyperparameters impact the operation of the system. We found a reference set of parameters that allowed us to miminize the execution time and energy expenditure while maintaining the highest accuracy we were able to obtain. We note that we were able to further reduce the processing time, even by reducing more the input size or sparsity of the signal. We suspect this is because, at this level of sparsity, the execution time is lower-bounded by the general operation time of the chip.

\subsection{Comparison between non-neuromorphic and neuromorphic models} \label{sec:CVK_Loihi_payload}

\begin{figure}[ht]
\centering
\includegraphics[width=\linewidth]{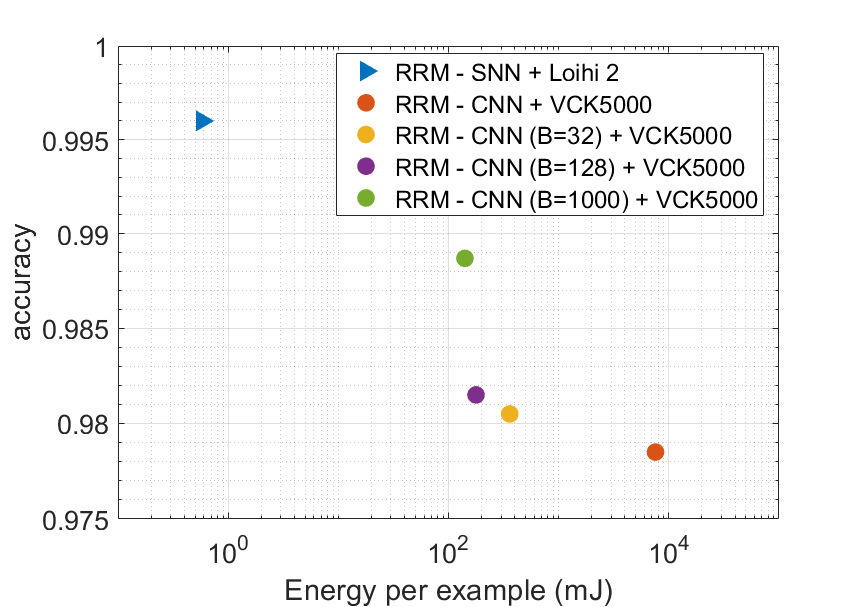}
\caption{Accuracy vs. Energy Consumption. Comparison between the execution of an SNN on Loihi 2 and the execution of a CNN on the CPU of the VCK 5000 chip.}\label{fig:vsLoihiAcc}
\end{figure}

\begin{figure}[ht]
\centering
\includegraphics[width=\linewidth]{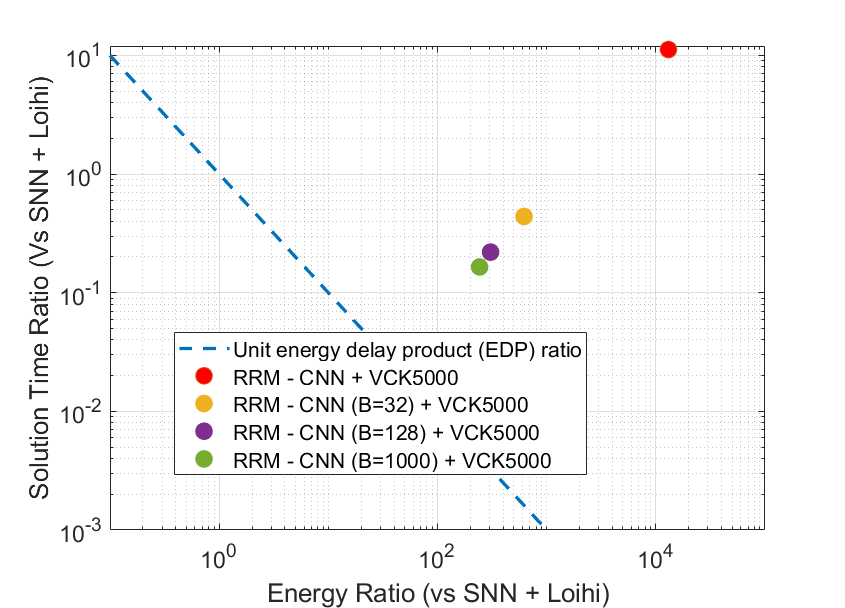}
\caption{Solution Time Ratio of a CNN on VCK5000 compared to an SNN on Loihi 2 versus the Energy Ratio of a CNN on VCK5000 compared to an SNN on Loihi 2. }\label{fig:vsLoihi}
\end{figure}

We present a comparative analysis between the results obtained from inference using SNN on Loihi 2 and the VCK5000 chip for the CNN model. 
Fig.~\ref{fig:vsLoihiAcc} presents a comprehensive comparison of accuracy and power consumption for both combinations. The results clearly demonstrate that the neuromorphic model consistently outperforms the non-neuromorphic approach in terms of accuracy while drastically reducing energy consumption. In fact, the energy consumption is up to two orders of magnitude lower than the models running on the VCK5000 chip.

One of the factors contributing to the superior performance of the neuromorphic approach is the deployment of the CNN model on the VCK5000 chip. To facilitate this deployment, the CNN model must undergo quantization, which involves reducing its accuracy. This quantization process inevitably leads to a loss of accuracy compared to the original values obtained in Section VI.A. Consequently, the lower accuracy of the CNN model quantized on the VCK5000 chip is one of the reasons for the difference in performance between the two approaches.

However, the primary and more significant reason behind the higher accuracy achieved through inference using SNN on Loihi 2 can be attributed to the characteristics of the traffic matrix $\bm{R}_{\tau}$. We have observed that the traffic matrix exhibits a large spatial sparsity, where a substantial portion of the values are zeros. The neuromorphic approach, particularly the SNN on Loihi 2, benefits significantly from this spatial sparsity, making it inherently well-suited to leverage such data characteristics.

The ability of the neuromorphic approach to process sparse data efficiently is a key advantage that enables it to outperform the non-neuromorphic approach. The unique design of the Loihi 2 chip allows for specialized processing of sparse data patterns, leading to optimized and accurate inferences. As a result, the neuromorphic models excel in extracting valuable information from the sparse traffic matrix $\bm{R}_{\tau}$, enabling them to achieve higher accuracy levels compared to the non-neuromorphic approach.

In Fig.~\ref{fig:vsLoihi}, we present the results of the comparative benchmarks for latency and energy in a unified view. This two-dimensional plot highlights the key advantages offered by neuromorphic hardware, such as the Loihi chip, as compared to commercially available programmable architectures. The dashed diagonal line represents the energy-delay ratio parity line~\cite{DaviesvsLoihi}, with benchmark points located below and to the left of this line indicating architectures that outperform Loihi. In contrast, points located above and to the right indicate superior performance for Loihi. In all the scenarios presented here, the conventional models lie above the parity line, indicating that neuromorphic algorithms running on dedicated hardware provide clear benefits to perform RRM on board of satellites.

\section{CONCLUSION}
This article presents an extensive investigation into the benefits of incorporating neuromorphic computing and SNNs for on-board radio resource management in SatCom systems. By leveraging innovative approaches, we addressed the challenge of implementing on-board RRM, comparing the performance of the proposed neuromorphic computing approach with a traditional CNN model. Our experiments demonstrate that SNNs, enabled by dedicated hardware, offer higher accuracy and significantly reduce energy consumption and latency. These remarkable results underscore the potential of neuromorphic computing and SNNs in improving RRM for SatCom, leading to better efficiency and sustainability for future SatCom systems.

To advance this research further, several avenues of investigation remain open. An important aspect is the implementation of the proposed approach in a real system, taking into account factors such as radiation tolerance, which holds great significance in the space environment. Moreover, future research could focus on optimizing the SNN architecture to achieve better performance and energy efficiency, considering the specific requirements and constraints of SatCom systems.

Another promising direction is the integration of continuous learning capabilities into the SNN-based RRM approach. This would enable the system to adapt and evolve over time, accommodating dynamic changes in traffic patterns and resource demands. By incorporating continuous learning, the system would enhance its adaptability and responsiveness in real-world scenarios, leading to more efficient resource allocation and management.

The findings of this study lay a solid foundation for the application of neuromorphic computing and SNNs in the field of SatCom RRM. Future investigations can build upon this work to further advance the state-of-the-art in SatCom systems, leveraging the benefits and insights gained from this comprehensive study.

\section*{ACKNOWLEDGMENT}
This work has been supported by the European Space Agency (ESA) funded under Contract No. 4000137378/22/UK/ND - The Application of Neuromorphic Processors to Satcom Applications. Please note that the views of the authors of this paper do not necessarily reflect the views of ESA. Furthermore, this work was partially supported by the Luxembourg National Research Fund (FNR) under the project SmartSpace (C21/IS/16193290). The work of O. Simeone was also supported by the European Union’s Horizon Europe project CENTRIC (101096379), by an Open Fellowship of the EPSRC (EP/W024101/1), and by Project REASON, a UK Government funded project under the Future Open Networks Research Challenge (FONRC) sponsored by the Department of Science Innovation and Technology (DSIT). The work of B. Rajendran was also supported  by the European Union’s Horizon Europe project CENTRIC (101096379) and an Open Fellowship of the EPSRC (EP/X011356/1). The authors gratefully acknowledge the support of Intel Labs through the Intel Neuromorphic Research Community (INRC) and Tomas Navarro as ESA officer. 

\bibliographystyle{ieeetr}
\bibliography{references.bib}

\end{document}